\documentclass[article,onecolumn,preprintnumbers,amsmath,amssymb]{revtex4}
\usepackage[dvips]{graphicx}
\usepackage[dvips]{color}

\usepackage{xcolor}
\definecolor{scwColorEdting}{RGB}{254, 1, 1}
\begin{document}

\title{Nematic and smectic liquid crystals modeling with stiff and free-joint Lennard-Jones chain molecules}
%% \title{Orientation ordering in mesogens as an effect of backbone alignment}
%Geometric Effects of Imperfect Orientation Ordering of Condensed Anisotropic Molecules}

\author{Wen-Jong Ma$^{1,2}$}
\email{mwj@nccu.edu.tw}
\author{Leng-Wei Huang$^{2,3}$}
\author{Shih-Chieh Wang$^{2,4}$}
\author{J\'an Bu\v{s}a Jr.$^{2,5,6}$}
\author{Natalia Tomasovicova$^{6}$}
\author{Milan Timko$^{6}$}
\author{Peter Kopcansky$^{6}$}
\author{Katarina Siposova$^{6}$}
\author{Chin-Kun Hu$^{2,4}$}
\email{huck@phys.sinica.edu.tw}

%\vskip 3 mm

\affiliation{ }
\affiliation{$^{1}$Graduate Institute of Applied Physics, National Chengchi University, Taipei 11605, Taiwan}
\affiliation{$^{2}$Institute of Physics of Academia Sinica, Taipei 11529, Taiwan}
\affiliation{$^{3}$Department of Physics, National Taiwan Normal University, Taipei 10610, Taiwan}
\affiliation{$^{4}$Department of Physics, National Dong-Hwa University, Hualian 97401, Taiwan}
%\affiliation{$^{4}$RIKEN Center for Computational Science, RIKEN, Chuo-ku, Kobe, Hyogo, Japan}
\affiliation{$^{5}$Laboratory of Information Technologies, Joint Institute for Nuclear Research, %Joliot-Curie 6,141980
Dubna 141980, Moscow Region, Russian Federation}
\affiliation{$^{6}$Institute of Experimental Physics, Slovak Academy of Sciences, Kosice 04001, Slovakia}

\date{\today}

\begin{abstract}
Liquid crystals (LCs) composed of mesogens play important roles in various scientific and engineering problems. How a system with
many mesogens can enter a LC state is an interesting and important problem. Using stiff and free-joint Lennard-Jones chain
molecules as mesogens, we study the conditions under which the mesogens can enter various LC phases.  The guideline is to eliminate
 the unwanted translational orders under a controlled fine-tuning procedure across a sequence of systems. Instead of monitoring
 the growth of order out of the disorder, we prepare a configuration of high orientation ordering and find out where it relaxes to.
 Such a procedure begins with a reference system, consisting of short chains of homogeneous soft spheres, in a liquid-vapor
 coexistence situation, at which the thermodynamic instability triggers a fast spontaneous growing process. By applying a short
 pulse of auxiliary field to align the dispersedly oriented clusters, followed by reducing the volume and, finally, changing
 the homogeneous molecules into heterogeneous chains, we are able to obtain a range of systems, including nematic and smectic LCs,
 at their stable ordered states. The model can be extended to study the influence of nanoparticles or external field on the LC structure.
\end{abstract}

\maketitle
\medskip

\vskip 3 mm

%\cite{}
%\pacs{05.50.+q, 05.70.Jk,05.10.Ln, 05.10.Cc}

%% \cite{71Stanley,14cjpHu,95jpa3dIsing,96jpaIsing,review,2012jcp}
%%  \bibitem{45LG}  Guggenheim, E. A. The Principle of Corresponding States. {\it J. Chem. Phys.} {\bf 13}, 253 (1945).

Simple matter composed of atoms (e.g. Helium, Copper, etc.) or molecules (e.g. oxygen, carbon dioxide, etc.)
with a small number of atoms in each molecule
can show phases of gas, liquid and solid which can be either crystal or amorphous solid.
The molecules in a liquid are disorder while the molecules in a crystal are arranged in a regular
three-dimensional array.
The phase transitions and critical phenomena of simple matter can be well described by some
popular models of statistical physics \cite{71Stanley,14cjpHu}.
For example,  the three-dimensional Ising model \cite{95jpa3dIsing,96jpaIsing} and the Lennard-Jones (L-J) particles system \cite{2012jcp}
can describe the liquid to gas critical phenomena \cite{45LG,review} very well \cite{14cjpHu,18Hu}.

The matter composed of chain molecules can show phases of liquid crystals (LCs) \cite{64SciAm-LC,NemaCholes,SmecColmn}
with some characteristics similar to liquids and some characteristics similar to crystals. There are many different phases of LCs.
In the nematic LC, the long axes of molecules are parallel to each other, but the molecules are not
arranged in layers, i.e. there is only orientational order and there is no translational order.
In the smectic LC, the molecules are arranged in layers and there are both orientational order (as in the nematic LC),
and translational order. In the smectic A LC, the molecules in a layer are at random;
In the smectic B LC, the molecules in a layer are in rows.
The presence of orientation ordered states in the "lyotropic" liquid crystal
can be theoretically understood as the result of the volume effect caused by the anisotropic core of the
mesogen.
Calculations based on hard model molecules showed the emerging of mesomorphic phases,
intermediate between isotropic fluid and crystal phase, on increasing density, where temperature plays no role.
In the "thermotropic" liquid crystals, on the other hand, the temperature-dependence of the phases
is associated with inter-mesogen attractive interactions. In the isotropic fluid regime, such interactions
suppress the entropic effect and, at the lower density and the lower temperature, render the fluid separating
 into liquid and vapor when the number of particles in a volume is inadequate to support the configuration as a uniform fluid.

On the basis of our previous works
on molecular models of polymers \cite{MH1,MH2,MH3,MH4,MH6}, in this paper we will construct simple models of
chain molecules which can show various phases of LCs. Our models can be extended to study
the effects of nanoparticles or external fields on the structure of LCs.

Historically, the biomaterials were the first found to possess liquid crystal phases, where
molecules with chain structures act as building blocks \cite{NemaCholes}.
Ever since, various forms of ordered states have been revealed \cite{deGennes,SSRMP,Textures}, where the internal structure
of the molecules and the inter-molecule interactions are relevant.
For uniaxial molecules, if the orientation ordering is not accompanied by any translation orders, the system is
in a `nematic' state. The orientation ordering
can appear with the presence of translation orders. The simplest form of the latter is
the formation of equally-spaced lamellar, where
one dimensional translation order is present. Depending on the relationship between
the direction of the ordered state and the direction of the normal of the  lamellar, we
have a class of `smectic' states.
Since the systems are soft, the lamellar may be modulated by spatially varying
envelopes \cite{deGennes,Textures,SmecColmn}.
Experimentally, the manipulations of liquid crystal materials, in which the molecules are
diamagnetic \cite{deGennes},
are often achieved by applying a
magnetic field to align the dispersedly oriented domains of molecules, so that
a configuration with all these domains pointing in the direction of the field is obtained.
The orientation ordering introduces elastic energy, that
mesomorphic phases with boundary effects are effectively described by continuum scenarios.
Recent experiments \cite{newLxtl} involve various effects contributed by doped particles of larger or comparable sizes
\cite{Slovakia1,Slovakia2}. The roles of geometries at interfaces and the presence of external fields, which may involve
doped particles, are less explored, especially theoretically.

Theoretical analysis suggests \cite{deGennes} that a large enough aspect ratio of molecules is necessary for
the formation of an orientation ordered configuration.
For molecules with weak geometric anisotropy \cite{MIVKB,ML,MT}, the fluctuation of the reorientation of molecules prevents
the system from having an orientation ordered configuration.
In a system of stiff polymer chains, the molecules have a strong tendency to
align with each other \cite{XtalpolyMelt,RMPpolymXtl,MH3,MH4,XtalPolym,MH6}, as soon as the
hindrance along the backbones of the molecules is sufficiently weak \cite{MH3,MH4,MH6}.
Under such a circumstance,
the formation of stacks of plate like crystallites \cite{RMPpolymXtl} on a substrate or that of
bundles \cite{MH3,MH4,MH6} in free space is guided by the backbone alignment.
The clustering process proceeds in such a way that the alignment among molecules
over local segments extends along the backbones, to reach a global extent \cite{MH3,MH4,MH6}.
It occurs at a lower density and at a temperature within the liquid-vapor two-phase
coexistence regime \cite{MH6}, in which the molecules have sufficient accessible volume for
reorienting to facilitate the alignment.
The clustering process proceeds spontaneously under the thermodynamic instability of that regime,
with the intermolecular attractive interaction \cite{Yukawattrac,GBattrac} as the driving force.
If the relaxation processes were carried out in an uniform configuration, outside the coexistence region,
at a higher density, it would require long time for the system to escape from those states of crowded
and possibly intertwined chain conformations, before it reached to have an ordered configuration.

In the present paper, the orientation ordering of mesogens is re-examined in the aspect of
backbone alignment, taking collections of short chains as our model systems.
By extending the studies of clustering
of polymer chains \cite{MH3}, we consider the alignment properties of shorter chains.
The model mesogens are designed
with built-in site-by-site heterogeneity along the chains. While the tendency of alignment is weakened by the
hindrance introduced by the heterogeneity and also by the enhanced reorientation fluctuations
accompanied by the reduced lengths of the chains,
we resume the tendency of alignment by applying a pulse of external field in the preparation of an initial configuration.
The true state, under the given thermodynamic conditions, is examined by finding the stability of
that configuration.
It is based on the simple assumption that the configuration of an orientation ordered equilibrium state or a stable state,
is readily accessed by a configuration with the orientations of the molecules
pointing in the same direction that the changes in the configurations
are achieved by some major shifting of molecules along the aligned direction. If not, we obtain an
orientation disordered configuration.

Chain-of-spheres based model mesogens with strictly rigid \cite{rigidchain} or semi-rigid structures
\cite{WA93,stifflex,DPDchain,warmlikeMD,helipart,coreflex}
have been studied by using Monte Carlo \cite{rigidchain, coreflex} or molecular dynamics simulation (MD)
\cite{WA93,stifflex,DPDchain,warmlikeMD,helipart} methods.
A soft anisotropic model molecule, such as the Gay-Berne molecule, can be considered as the  equivalence of the
coarse-graining \cite{GayBerne} of such a chain.
In this study, we employ the MD approach, beginning with systems of homogeneous free-joint Lennard-Jones (L-J) chains,
where the linking between a pair of nearest neighboring monomers along a chain is by a rigid
bond as in \cite{MH1,MH3}.  We assign L-J
interactions between non-nearest neighbors along a chain and between any pair
of monomers on two different molecules. With a short bond length \cite{MH3} (see Methods), we obtain models
of stiff polymers and systems of which undergo aggregation processes in the coexistence
regimes to form a major cluster with bundled structure \cite{MH3,MH6}.
We consider chains of lengths below the persistent length (see Methods).
The molecules are, therefore, nearly straight and retain the tendency to align with each other to form ordered domains in the
coexistence regimes.
The orientation ordered domains are found to have strong translation orders at the same time,  and a smectic liquid crystal
can thus formed. Based on such model molecules, we add one additional monomer on each end of a chain,
imposing the connection by a spring \cite{coreflex}, and vary
the sizes of the monomers along the chain.
In proper tuning the heterogeneity of the model chain, we can readily prepare a system in its orientation ordered state.
In this approach, we make use of the aspect of backbone alignment in systems of stiff polymer
chains in a coexistence situation,
in which the clustering processes occur spontaneously under the driving force of thermodynamically instability.
The introduction of heterogeneity to the model chain
effectively suppress or eliminate translation ordering, that is carried out in a well-tracked manner.
We show that, for a model chain of only
seven monomers, with an aspect ratio below 3.2, we are still
able to prepare a system in its orientation ordered states, with or without translation orders,
 i.e., we can prepare both smectic and nematic liquid crystals from linear L-J
molecules.

\vskip 4 mm
\noindent{\large \bf Results}

\vskip 2 mm
\textbf{Orientation ordering of short homogeneous chains.}
Figure~\ref{FIG1} shows the snapshot of a system of $N=400$ free-joint homo-chains, each of length $n=10$,
under relaxation in a thermodynamically
unstable, liquid-vapor
coexistence situation. The chain length $n=10$ is only one-tenth of
that for those chains in a system having the same interaction parameters, considered in
the study of polymer aggregation
\cite{MH3,MH4,MH6} and is less than the persistent length $l_p$
found for the
latter system,
where $l_p$ is greater than ten %approximately fifteen
times of the bond length $b$ (see Methods).
The configuration in Fig.~\ref{FIG1} is
at the late stage of a clustering process, which shows
the formation of several ordered domains.
The general scenario of clustering is similar to those described in ~\cite{MH3,MH4,MH6}.
We find the main body of the configuration consists of a layer of nearly straight, mutually parallel chains, which are
tilt from the normal of the plane on which the chains are in a crystalline structure (see Fig.~\ref{FIG1}(b)). Two smaller ordered
clusters attach to the layer, with their orientations in parallel to the plane.
The three clusters occupy approximately, in total, one-quarter of the box, leaving the rest volume empty.
To cut down the chain length further, we expect increased reorientation fluctuation in individual molecules.
The clusters develop orientation ordering globally only at higher densities,
with the help of the effect of close packing.
Figure~\ref{FIG2} shows that for a system
with $N=800$ homogeneous chains, each of length $n=5$, which is one-half of that for the system in Fig.~\ref{FIG1},
the ordered domain extended to fill the whole simulation box.

The orientation ordering of a system is quantified by the traceless tensor
\cite{deGennes}
\begin{equation}
Q_{\alpha \beta} \equiv
\frac{1}{N}
\sum_{i=1}^{N}
\left(\hat{n_i}\right)_{\alpha}
\left(\hat{n}_i\right)_{\beta}
-
\frac{1}{3}
\delta_{\alpha \beta}
\label{order},
\end{equation}
where $(\hat{n}_i)_{\alpha}$ is the $\alpha$-component of the direction vector $\hat{n}_i$ of the molecule $i$,
with $\alpha$, $\beta$=$x$, $y$, or $z$.
The $3\times3$ matrix with entries given by Eq.~(\ref{order})
is diagonalized to obtain the eigenvalues
$\frac{2}{3} S \ge \eta - \frac{1}{3} S \ge -\eta - \frac{1}{3} S$,  in descending order.
The value of $S$, in between zero and one, quantifies the orientation ordering of the system. For $S>0$,
the unit eigenvector $\hat{n}$ corresponding to the eigenvalue $\frac{2}{3} S$ specifies the overall direction of the ordering.
Correspondingly, we refine the analysis of orientation orders to include the information on intra-molecular ordering.
We compute the tensor
\begin{equation}
Q^{\rm bond}_{\alpha \beta} \equiv \frac{1}{N
\left(n-1\right)
}
 \sum_{k=1}^{N
\left(n-1\right)
 }
 \left( \hat{b}_k\right)_{\alpha}
 \left( \hat{b}_k\right)_{\beta}
- \frac{1}{3} \delta_{\alpha \beta}
,
\label{orderbond}
\end{equation}
to obtain eigenvalues $\frac{2}{3} S^{\rm bond} \ge \eta - \frac{1}{3} S^{\rm bond} \ge -\eta - \frac{1}{3} S^{\rm bond}$,
based on the direction vectors $\hat{b}_k$ of the individual bonds
$k=1, \dots, N(n-1)$.
For the configuration in Fig.~\ref{FIG1}, of the system with $N=400$ and $n=10$, where there are one major ordered
layer and two small pieces of ordered domains, we have
$S=0.817$ and
$\eta=0.0465$,
for molecules (Eq.~(\ref{order})). Those for bonds (Eq.~(\ref{orderbond})) are
$S^{\rm bond}=0.813$ and
$\eta^{\rm bond}=0.0462$.
For the configuration of the system with $N=800$ and $n=5$ in Fig.~\ref{FIG2}, the space is filled with only one major ordered domain,
the values for
$S=0.975$ and
$S^{\rm bond}=0.969$
are very close to unity.
The parameters
$\eta=6.94 \times 10^{-4}$ and
$\eta^{\rm bond}=6.34 \times 10^{-4}‬$
are quite small. The
orientations of the molecules as well as those of the bonds are very close to the overall direction of the ordering in this system.

The spatial variations of bond orientations and the orders in translations can be
obtained by calculating
radial distribution functions.
For orientation orders, we calculate the distribution of direction cosine as a function of site-site distance.
$g_{\rm cos}(r_{\rm site})$ describes
the mean direction cosine
$\hat{b}\cdot \hat{b'}$ as a function of the distance
$r_{\rm site}$  between the two sites where the bonds $\hat{b}$
and $\hat{b'}$ are at, respectively.
For translation orders, they are collected in accord with the directions of
 individual molecules.

 In Fig.~\ref{FIG2}, we demonstrate how such functions are calculated, by projecting the positions of the centers
of the chains, either on the plane perpendicular to the reference chain (Fig.~\ref{FIG2} (a)) or on the line along the direction
of that chain (Fig.~\ref{FIG2} (b)). Let $V$ be the volume occupied by $N$ mesogens and $\vec{r}_{ji}$ be the position vector from
the $i$-th mesogen to the $j$-th mesogen. The two-dimensional and the one-dimensional radial distribution functions,
$g_{\bot}(r_{\bot})$ and $g_{\|}(r_{\|})$, given by
\begin{equation}
2 \pi r_{\bot} h g_{\bot} \left(r_{\bot}\right) \Delta r_{\bot}
=
\frac{1}{N}
\sum_{r_{\bot} < R \le r_{\bot}+\Delta r_{\bot}}
\sum_{i=1}^{N}
\frac{V}{N}
\sum_{j=1}^N
\left(
	\delta
	\left(
		\left|
			\vec{r}_{ji}-
			\left(\vec{r}_{ji} \cdot \hat{n}_{i}	\right)
			\hat{n}_{i}
			\right|
		-R
		\right)
	\Theta \left(\frac{h}{2}-\vec{r}_{ji} \cdot \hat{n}_{i}\right)
     \Theta \left(\frac{h}{2}+\vec{r}_{ji} \cdot \hat{n}_{i}\right)
     \right),
\label{gperpen}
\end{equation}
and
\begin{equation}
 \pi \left(\frac{r_s}{2}\right)^2 g_{\|} \left(r_{\|}\right)  \Delta r_{\|}
 =
 \frac{1}{2N}
 \sum_{r_{\|} < R \le r_{\|}+\Delta r_{\|}}
 \sum_{i=1}^{N}
 \frac{V}{N}
 \sum_{j=1}^N
 \left(
	\delta \left( |\vec{r}_{ji} \cdot \hat{n}_{i}|-R\right)
	\Theta \left(r_s
	 - \left|\vec{r}_{ji}-
	    \left(\vec{r}_{ji} \cdot \hat{n}_{i}\right)
	    \hat{n}_{i}
	   \right|
	   \right)
    \right)
,
\label{gpara}
\end{equation}
respectively, for $0 \le r_{\bot}, r_{\|} \le \frac{L}{2}$, provide the information of the translation orders, with the coarsening scales of $h$ (the height of a cylindrical shell) and $r_s$ (the radius of a circular plate), respectively,  in a simulation box of linear size $L$.
We will
consider these three functions for the final configuration (Fig.~\ref{FIG2}) as well as for typical instantaneous snapshots over the quenching process.

Now, we go into the details of the processes in obtaining the ordered configurations shown in Figs.~\ref{FIG1} and~\ref{FIG2}.
We focus on the case of Fig.~\ref{FIG2}, which is taken as a reference system, based on which we construct models of liquid crystal.
The configuration is reached via an isobaric quenching process (see Methods for technical details)
from vapor.
Figure~\ref{FIG3} shows data of instantaneous temperature $T^*$, instantaneous pressure $P^*$, density $\rho^*$ and the order parameter
$S$ (Eq.~(\ref{order})), plotted by pairs in six plots (a quantity with a superscript $^*$ is in the reduced unit, defined in Methods). The quenching process is shown by plotting the
 time-averages (red line) in Fig. 3(c), 3(d), and 3(f).
%%  $T^*$ versus $\rho^*$ (center), $P^*$ versus $ \rho^*$ (bottom) and $T^*$ versus $P^*$ (right).
Under the externally imposed controlling pressure, at a preset value $P_0=0.1$ (in unit of $\epsilon \sigma^{-3}$, see Methods),
the system
decreases its volume at the lowered temperature, to compensate the internal pressure which lost its contribution from the kinetic energy.
The compensation is also from the clustering of the chains, which increases the internal stress.
From the plots, there is an apparent abrupt change in the amplitude of fluctuation in $P^*$ and, simultaneously, in $\rho^*$,
corresponding to the crossover from vapor to liquid. From the
plot of  $P^*$ versus $\rho^*$ and that of  $T^*$ versus $P^*$, we see that,
the system is maintained in or close to thermodynamic equilibrium, where the time-averaged
instantaneous pressure (red line) is
in agreement with the target pressure ($P_0$ = 0.1), on both the vapor and the liquid sides
(see Fig.~\ref{FIG3}(f)).  The process is close to the boundary of the coexistence
region (see Fig.~\ref{FIG3}(c)) in the liquid regime, before it comes to the
next occasion of abrupt change, in the values of $\rho^*$ and  $S$.
It signals the crossover from liquid to crystal.
The time-averaged instantaneous pressure (red lines in Fig.~\ref{FIG3}(d) and Fig.~\ref{FIG3}(f)),
is apparently smaller than the
target value ($P_0$ = 0.1) at this crossover and is even negative on entering the crystal regime.
The system is therefore in the non-equilibrium situation in entering
the latter regime, under the fast quenching performed in the simulation. Over the crystal regime,
the changes in density and the order parameter are quite small in comparing with the corresponding changes in
pressure and temperature.
In contrast to the large change in $S$ at the crossover from liquid to crystal (in plots (a), (b), and (e)),
no corresponding change can be recognized in the value of $S$ at the crossover from vapor to liquid.

In Figs.~\ref{FIG4}, we plot the distributions $g_{\bot}(r_{\bot})$ (Eq.~(\ref{gperpen})), $g_{\|}(r_{\|})$ (Eq.~(\ref{gpara})) and
$ g_{\rm cos}(r_{\rm site})$,
for the ordered configuration Fig.~\ref{FIG2},
as well as those instantaneous spots before and after the crossover to each of the two crossovers.
The curves are labelled by the same symbols that their corresponding spots in Fig.~\ref{FIG3} are marked.
Note that, while the heights of the peaks and the depths of the valleys of
$g_{\bot}(r_{\bot})$  depend on the height $h$ of the cylinder, the chains with their centers
within which are projected (Fig.~\ref{FIG2} (a)), so do the locations of the peaks and the valleys. The same is true
for $g_{\|}(r_{\|})$, that the dependence of the sizes and the locations of those peaks and valleys on
the choice of the radius $r_s$ of the cylinder described in Fig.~\ref{FIG2} (b).
For the systems considered in this study, we observe no dependence of the locations of the peaks and the valleys
in either $g_{\bot}(r_{\bot})$ or $g_{\|}(r_{\|})$,  on the choice of $h$ and $r_s$, respectively.
The results indicate no twists in the structures are present.
The curves in Fig.~\ref{FIG4}, as well as those for other systems considered in this paper,
are for $h=r_s=\frac{L}{2}$, where $L$ is the length of each edge of the simulation
box.

In Fig.~\ref{FIG4}(c), $ g_{\rm cos}(r_{\rm site})$  is maintained at
high values over the whole range of $r_{\rm site}$ considered in the figure, once the system enters the crystal
regime,
since there is only one domain in the system
and the chains are highly orientation ordered.
There is almost no change in the values of
$ g_{\rm cos}(r_{\rm site})$, in the range
beyond $r_{\rm site}$=1.2($\sigma$), when the system moves from the spot
on entering the regime (marked by diamond), to that of the end of simulation (marked by circle), even though the system is
in non-equilibrium, with negative pressure and with changing density and temperature (Fig.~\ref{FIG3}).
In Fig.~\ref{FIG4}, we see the agreement between the locations of the peaks in $g_{\bot}(r_{\bot})$,
and those contributed by a triangular lattice, taking the location of the highest peak as the edge length.
There is, however,
an additional peak in between $r_{\rm site}=0$ and the location of the highest peak. The presence of such a peak indicates
that the chains point in
a common direction tilt from the normal of the stacks of planes of triangular lattice. The configuration,
therefore, has a tilt smectic structure \cite{NemaCholes,SSRMP,Textures}.

For the two systems of homogeneous chains considered so far, the orientation orderings are always accompanied by
translation orders (smectic liquid crystal).
It is found that, in the present of orientation ordering, the translation orders can not be removed completely
%when we change
by changing the thermodynamic conditions of the systems.
 %, especially, near the liquid-vapor coexistence regime.
In the following, we explore and design model systems in an attempt to
remove, either partially or totally the translation orders, in a well-tracked manner.

%% ~\\
%% \textbf{Pathways across interaction-parameters to reach orientation ordered or disordered states}
%% ~\\

\vskip 2 mm

{\noindent \bf Reaching orientation ordered or disordered states via changing interaction-length parameters.}
In building model mesogens \cite{stifflex,slowtransLxtl,fusedhard,phasdiag0,phasdiag1,phasdiag2,hardspher,SimuLxtl,Potsemiflx},
the presence of orientation ordered phases and the transitions among these phases and
the disordered phase can be realized by grasping and quantifying a few key parameters
in molecular geometry
and in inter-molecular interactions. While it is feasible to manipulate these parameters to explore various qualitative possibilities
in the phase diagrams \cite{phasdiag0,phasdiag1,phasdiag2}, it is also intended to reproduce
specific experimental data quantitatively.
On the experimental sides \cite{NemaCholes,classHird,classGMDZC}, it is often to address
the issues by tracing the individual species of molecules, on how the thermodynamic properties are changed
with the modifying on the effect of excluded volume \cite{classHird,classGMDZC}. Similar task is also employed and extended in the
model of soft anisotropic molecules \cite{GayBerne}, where the tuning of the shape as well as the interaction
are manipulated to generate various possible topographies of phase diagram \cite{GBattrac,phasdiag0,phasdiag1,phasdiag2}.
In this study, instead of exploring the whole phase diagram, we inquire how the translation order of
the model systems can be diminished in a well-tracked manner. In particular,
we investigate how the translation ordering is eliminated partially or completely in
systems of molecules with a small aspect ratio \cite{GBattrac,stifflex,hardspher},
at lower temperatures, near the liquid-vapor coexistence \cite{GBattrac} regimes.
Based on the observations on the ordering, within or in the
vicinity, of the
two-phase coexistence regimes (Figs.~\ref{FIG1}-\ref{FIG4}),
the issue is rephrased by raising the question: how a nematic ordering can be developed
effectively between the two extreme situations,
one with the backbone dominated clustering occurring in a rarefied density and
the other with close-packed ordering emerging in supercooled dense fluids.
In preparing a system at an orientation ordered state,
we adapt a route taking advantage of the former effect. Instead of obtaining it from the relaxation of an isotropic configuration,
we start with an orientation-aligned configuration. Here, we build a model chain
of $n=7$ monomers \cite{WA93}, with five monomers connected sequentially by rigid bonds as a core,
and two additional monomers, each connected by a spring to one of the two end monomers of the core \cite{stifflex,coreflex}.
Both model studies \cite{classHird,stifflex} and experimental analysis
\cite{classHird,classGMDZC} suggested
that the flexible tails play the role in softening the strict orders in translation, to allow for the emerging of purely
orientation (nematic) ordering in the states between the isotropic fluid and the crystal in the phase diagram.
In our models,
all monomers on a chain are free jointed. Each monomer interacts with any other monomer
on the same chain, except for its nearest neighbor(s),
and every monomer on another chain, via L-J potentials. The bond lengths are the same
as those in Fig.~\ref{FIG1} and in Fig.~\ref{FIG2}.
Under a given thermodynamic condition, we are allowed to change the size and the strength parameters of the L-J potentials of
the seven monomers along the chain to obtain a range of systems, each with specified status in its
ordering property. To render such changes in the interaction parameters in a well-tracked manner,
we use the same strength parameter for all
monomers along the chain
and keep it fixed. With symmetrically arranged monomers along the chain (see text below), we change only two size-parameters.

%Specifically, to have an orientation ordered state, the suppression or the elimination of translation order
%is well-tracked.

We start with system M, of semi-homogeneous chains, each of which has identical L-J parameters
for all the seven monomers (see Methods  and Table~\ref{Interaction}). Each chain in the %model
system has a five-monomer core which is
equivalent to those %a five-monomer chain, with a rigid bond between each pair of nearest neighbors,
in the system that has been described by Figs.~\ref{FIG2}-\ref{FIG3}. The latter system, composed of $N=800$
chains, becomes ordered under a
quenching process, at the final stage of which a major domain
of smectic-B crystal forms (Figs.~\ref{FIG2} and \ref{FIG4}), which have six-fold symmetry in each layer.
In system M, with the extra monomers connected to the two ends of the core, %a single chain
by springs, the tendency of alignment between the molecules %in system M
is weaker than that for the system in Fig.~\ref{FIG2}, of molecules without such monomers.
%less perfect. %has a weakened tendency to extending its local translation ordering
The system has, therefore, a weakened tendency to develope global translation (crystal) orders when it is quenched.  We take system M
as a starting point for further fine-tuning, in order to diminish the translation orders in a well-tracked manner. To reduce the possible
artifact caused by the simulation box, we maintain the box to be a cube to avoid a biased direction and choose the number of chains
that is sufficient to produce
one well-recognized domain of orientation ordered configuration.
%have the less tendency to form a single domain of crystal when the system is quenched.
We found a system composed of $N=2000$ of
such seven-monomer chains is appropriate.

Fig.~\ref{FIG5} summarizes how the procedures of preparation are carried out.
The system is put in a large volume and quenched within the liquid-vapor coexistence regime.
It undergoes a clustering process.
From the two cases, Fig.~\ref{FIG1} for $n=10$ and
Fig.~\ref{FIG2} for $n=5$, it is not surprising to observe the formation of
ordered, oriented dispersedly, flat clusters.
The scenario is similar, in part, to the clustering in an isotropic fluid of
hard spherocylinders \cite{hardsphercluster}. Since the latter molecules
have no attractive inter-molecule interaction, there is only one phase in fluid.
In our case, we intend to have models possessing typical liquid crystal phases.
A preliminary simulation shows that it is inefficient to carry out in system M, the
whole quenching process corresponding to the one described in Fig.~\ref{FIG3}.
To avoid the time-consuming process of spontaneous colascence and to leave rooms for improving
for classes of liquid crystal models, we carry out the simulations in the opposite direction
to it is in the conventional approach.
Instead of monitoring the growth of order out of the disorder, we prepare a configuration
of high orientation ordering and find out where it relax to.
For system M, we align the domains of small clusters
by applying an auxiliary external field (see Methods). It is to
revive the tendency of backbone alignment, which has been lost in cutting down the lengths of the chains.
We go over, then, a sequence of systems in tuning
the interaction parameters of the molecules, the system of which may or may not remain in an ordered configuration.
The reaching of the objective final state is
carried out as a relaxation from that configuration.
The schematic chart in Fig.~\ref{FIG5}, which contains four planes in each of which we include a plot
of temperature versus density,
to illustrate how such a procedure is realized in simulation, to produce a system of
heterogeneous chains in its ordered or disordered state.

In the beginning, we have a randomly prepared non-uniform configuration
in non-equilibrium coexistence situation for system M
(a typical snapshot, marked by M$_1$ in Fig.~\ref{FIG5}).
%Based on the data plotted in Fig.~\ref{FIG3},
%we estimate \cite{phasdiagalkane,phasdiagalkane2} the location of the coexistence
%region for system M, in Fig.~\ref{FIG5}.
In an isobaric quenching process, the volume of system is reduced. The process is
in a controlled non-equilibrium relaxation with the volume of
the system changing periodically, enveloped by a decaying oscillating amplitude, as an effect of the internal spring potentials
at the end monomers of the chains. Dispersedly oriented local domains are generated. In
applying a pulsed external field (see Methods) to align
the orientation of the seven-monomer semi-homogeneous chains in accord with the direction of
the field (configuration M$_2$). The orientation ordering
is retained even after the field is turned off (configuration M$_3$).
While the system is still in a non-equilibrium coexistence situation, we reduce its volume to eliminate the void space
with a high pressure.
Before we proceed to carry out the tuning of interaction parameters, the configurations (M$_{\rm I}$ and M$_{\rm II}$)
of the systems are in six layers with modulations deviated from planar structures.
The parameter $S$ over time is maintained, on the
same level as that before turning off the field, with the values above 0.9.
Based on the path of quenching plotted in Fig.~\ref{FIG3}, for the system with $n=5$,
we are sure the above process for system M is within %\cite{phasdiagalkane,phasdiagalkane2} % the location of
the coexistence
region. % for system M, in Fig.~\ref{FIG5}. %which is plotted the region bounded by gray dashed line in Fig.~\ref{FIG5}.
Indeed, the boundary of coexistence region shifts outwards
if the number of monomers  of each molecule is increased \cite{phasdiag1,phasdiagalkane,phasdiagalkane2}, or if the
anisotropy of individual molecules is enhanced \cite{GBattrac} in case that the heterogeneity along the
backbone is introduced.

We, then, starting from the configurations M$_{\rm I}$ and M$_{\rm II}$, resepctively, at densities $\rho_{\rm I}$ and $\rho_{\rm II}$,
adjust the interaction parameters by consecutive fine-tunings,
to reach new states with the translation orders eliminated totally or partially.
The procedures are carried out adiabatically with the volume fixed. To keep the systems well-tracked on the changes, we maintain all systems
at the intermediate steps at the same
instantaneous temperature $T_0$, which is achieved by removing (or adding) heat repeatedly.
In between two consecutive heat-adjusting steps,
the system is adiabatically evolving.
The two routes, starting from M$_{\rm I}$ and M$_{\rm II}$, respectively, are
named as pathway I and pathway II. The states on each pathway have a fixed density and a fixed controlling temperature.

  Points $(\rho_{\rm I}, T_0)_{\rm A}$
and $(\rho_{\rm I}, T_0)_{\rm C}$ in Fig.~\ref{FIG5}, specifically, %on plane C
are reached from the configuration taken at the point $M_{\rm I}$ ($\rho^*=\rho_{\rm I}$, $T^*=3.34985$)
on plane M via the pathways I, crossing the space of
interaction parameters, keeping the reduced density and the controlled temperature
fixed at $\rho_{\rm I}$ and $T_0$, respectively.
With the same controlled temperature and a different choice of reduced density, $\rho_{\rm II}$,
we defines pathways II, via which point $(\rho_{\rm II}, T_0)_{\rm B}$ on plane B is
reached from the configuration taken at the point $M_{\rm II}$ ($\rho^*=\rho_{\rm II}$, $T^*=3.31122$) on plane M.
The orientation-ordered (in $z$-direction), zig-zag layered configurations for the pathways I and II are
obtained on the quenching, first by applying a pulse of applied auxiliary field in $z$-direction, in between the spots labeled by
``field on" and ``field off", respectively,
to align the
dispersedly oriented ordered domains in the partially filled simulation box, % (snapshosts $M_1$ and $M_2$).
followed by imposing an external %the controlled value of
pressure $P_0=20.0$ (starting at  the spot marked by ``pressurizing on") at the controlled temperature $T^*=T_0$,
to allow the space to
be filled fully by the molecules. %}
%we focus on three typical states reached by the procedures. Each of the three systems A, B and C,
%has its own set of interaction parameters (see Table~\ref{Interaction}).
The configuration shown at
point $(\rho_{\rm I}, T_0)_{\rm A}$ on the plane (system) A, is obtained by a sufficient relaxation after the reaching
of the point by the sequence of adjustments of the parameters over pathway I. It shows the typical features of a smectic
state, with a six-layer lamellar structure. The configuration at point $(\rho_{\rm I}, T_0)_{\rm C}$, reached by the same pathway,
is that of an isotropic fluid, where there is no any orientation or translation order.
The relaxed configuration at point $(\rho_{\rm II}, T_0)_{\rm B}$ on plane B, possess a  good
orientation order and no any translation order.
The system is in a nematic state at that point.

In Fig.~\ref{FIG6}, we quantify the orderings for each of these three configurations.
For comparison, we also put the data of an ordered
configuration, plotted in Fig.~\ref{FIG4}, for the system of $n=5$ chains. It is apparent that the radial distribution $g_{\|}(r_{\|})$ for
the configuration at $(\rho_{\rm II}, T_0)_{\rm B}$ shows no sign of ordering at all, in spite of strong long range
angle correlation in the function $ g_{\rm cos}(r_{\rm site})$. The configuration,
is, indeed, in a nematic state.
We also carry out the same comparison that has been done in Fig.~\ref{FIG4}, for the configuration at
 $(\rho_{\rm I}, T_0)_{\rm A}$, between the locations
of the peaks in $g_{\bot}(r_{\bot})$ with those from the expected
triangular lattice. It turns out, while
the oscillations in $g_{\|}(r_{\|})$ suggest good one-dimension translation orders, the triangular ordering is restricted
within a shorter range.

At this point, it is important to verify that the ordered configurations reached at point
$(\rho_{\rm I},T_0)_{\rm A}$ and point $(\rho_{\rm I},T_0)_{\rm B}$, respectively,
by the pathways I and II described in Fig.~\ref{FIG5}, are not transient.
It is straightforward
to carry out the tests by heating up
the systems. In fast heating, both configurations reached in Fig.~\ref{FIG5}
are stable, until the instantaneous temperatures coming up to $T^* \approx 5.0$,
beyond which we observe melted isotropic fluid configurations.

%% ~\\
%% \textbf{Stability in orientation ordered configuration prepared by applying auxiliary field}
%% ~\\

\vskip 2 mm

{\noindent \bf Stability in orientation ordered configuration prepared by applying an auxiliary field.}
We carry out a few procedures starting from configurations other than those in system M,
to reach each of the two points, $(\rho_{\rm I},T_0)_{\rm A}$ and $(\rho_{\rm I},T_0)_{\rm B}$,
across the parameter space.
It is verified
that the reached states in system A and system B
share, respectively, the same features as those obtained in Fig.~\ref{FIG5},
after, in each case, applying a pulse of field to an isotropic configuration.

In the first test for the nematic configuration at $(\rho_{\rm II},T_0)_{\rm B}$ in system B, we %carry out the first test
%by starting
start from a stable isotropic configuration at the point $(\rho_{\rm I},T_0)_{\rm C}$,
moving reversely along pathway I,
which has been defined as
the pathway with fixed density $\rho_{\rm I}$ and controlling temperature $T_0$, to reach
point $(\rho_{\rm I},T_0)_{\rm B}$, the intersection of pathway I on plane B.
The density is then increased by pressurizing the system, to reach the vicinity of $(\rho_{\rm II},T_0)_{\rm B}$ on plane B:
\[
(\rho_{\rm I},T_0)_{\rm C} \rightarrow (\rho_{\rm I},T_0)_{\rm B} \rightarrow (\rho_{\rm II},T_0)_{\rm B}.
\]
In the second
test, a partially melted configuration
at the point $(\rho_{\rm I},T_0)_{\rm B}$ obtained in the middle of the original procedure,
is pressurized on plane B, to reach  the vicinity of $(\rho_{\rm II},T_0)_{\rm B}$:
\[
{\rm M}_{\rm I} \rightarrow (\rho_{\rm I},T_0)_{\rm A} \rightarrow (\rho_{\rm I},T_0)_{\rm B} \rightarrow (\rho_{\rm II},T_0)_{\rm B}.
\]
After applying a pulse of external field in either case, the system relax to
the stable nematic state. In Fig.~\ref{FIG6}, we observe that the pair distributions (dotted curves) of the relaxed
configuration obtained in the first test, are in agreement with
those counterparts for the configuration
in the original procedure (Fig.~\ref{FIG5}).

To test for the smectic configuration at $(\rho_{\rm I},T_0)_{\rm A}$
in system A, we start from a melted isotropic configuration, obtained by heating-up, controlled at
$(\rho_{\rm II},T_0)_{\rm B}$ of plane B and
carry out a revert
process via pathway II to $(\rho_{\rm II},T_0)_{\rm A}$ of system A, followed by a depressurizing process on plane A
to reach the vicinity of $(\rho_{\rm I},T_0)_{\rm A}$:
\[
(\rho_{\rm II},T_0)_{\rm B} \rightarrow (\rho_{\rm II},T_0)_{\rm A} \rightarrow (\rho_{\rm I},T_0)_{\rm A}.
\]
In relaxation from a nematic-like configuration (Fig.~\ref{FIG7}(a)),
obtained by applying an pulsed field,
the system starts to develop layers (Fig.~\ref{FIG7}(b)), toward a configuration with
the well recognized lamellar structure (Fig.~\ref{FIG7}(c)),
which is not identical to the configuration shown in Fig.~\ref{FIG5}.
It is, nevertheless, sufficient to identify that point $(\rho_{\rm I},T_0)_{\rm A}$ and its vicinity in system A,
be smectic in nature \cite{deGennes,Textures,SmecColmn}.

On the observation that the density $\rho_{\rm II}$ is greater than $\rho_{\rm I}$ and, at the latter density, system A is smectic,
it is sensible to raise the question : could it be
that the true state for system B at $(\rho_{\rm II},T_0)_{\rm B}$ were smectic, with layered structure,
and the nematic configurations obtained in simulation were in supercooled or glassy situations?
That is, the lack of layered ordering in the configurations
observed in simulations were
the consequence of retarded relaxations in an overly crowded space. If this were the case, the relieving of the crowdness
by reducing the density, say, to the point
% $(\rho_{\rm II},T_0)_{\rm B}$ should be well inside the territority of the smectic phase and
%the state at
$(\rho_{\rm I},T_0)_{\rm B}$, %which has a less crowded space,
would possibly lead to obtain smectic configurations. %would be smectic.
 Careful tests on the stability of ordered initial configurtions at $(\rho_{\rm I},T_0)_{\rm B}$ reveal that
all the configurations relax to lose the orientation orders as well as the translation orders, and to reach the configurations of an isotropic fluid.
It is unlikely that the true state %configurations
at $(\rho_{\rm II},T_0)_{\rm B}$ were smectic. %obtained in our simulations be supercooled or glassy.
The nematic nature at this point %$(\rho_{\rm II},T_0)_{\rm B}$
is, therefore, robust.

The testing procedures described above have shown that, by imposing pulsed auxiliary fields to a system
of anisotropic molecules at various initial configurations, we are able to find the true status of ordering for the system, under a
given thermodynamic condition. For the models considered in Fig.~\ref{FIG5}, two types of field-induced orientation orderings
have been prepared as the initial configurations.
One type is characterized by the presence of the lamellar structure. The other one has completely no translation orders in
each of its configurations.
For a given thermodynamic condition with fixed density and temperature, the true equilibrium or stable state is determined
by cross-examining the relaxation from each of those initial configurations to find out which characters,
among the smectic-like lamellar, the nematic-like purely orientation-ordered and the isotropic ones,
is present consistently in the final configuration.

~\\
\textbf{Discussion}
\vskip 2 mm

The tuning of the interaction parameters in Fig.~\ref{FIG5} is delicate and important. This is evidenced by the fact
that the density at the point of smectic state $(\rho_{\rm I}, T_0)_{\rm A}$ in system A
is non-negligibly smaller than that of nematic state at $(\rho_{\rm II}, T_0)_{\rm B}$ in system B.
Such a result excludes the attempt to take the volume packing effect
as the sole important factor, responsible for the presence of an ordered phase in a liquid crystal system,
according to which the system would be more favorable in a smectic state at the higher densities than in a nematic state is.
The current results suggest, instead, that the tiny changes in the interaction parameters from system A to system B,
which are only a few tenth-of-thousandth
part of unity (Table~\ref{Interaction}), are significant enough to dominate the difference
in symmetry, between the orderings at the two points, respectively, of the two systems.

The transitions between the isotropic fluid and the anisotropic liquid crystal phases
can be phenomenologically tracked by the temperature-dependence of the minimum in the
free-energy by taking the direction tensor of the whole system as the varying parameter.
%The fact that the free-energy has also the dependence on the spatial variations of the directions of the mesogens, in forms of
%elastic energy, refines the ordered phases.
%Therefore, various forms of spatial variations, combining
%the translation and the orientation orders,
%are possible.
In modern material and biophysical studies, the phenomenological descriptions involving interfacial properties with
confining boundaries \cite{nematicshell} or with doped particles \cite{Slovakia1,Slovakia2}, at length scales comparable with the sizes
or the characteristic lengths at molecule scales are
required to include effects at the mesoscopic scales.
The idea that the trend to reach an orientation ordering situation in a system of mesogens
is connected with the strong tendency of backbone alignment in systems of
stiff polymer chains, has been proven useful in that,
it inspires the designing of computation methods, to locate and to generate
in a model system, the intended
ordering properties. The approach emphasizes the capability of preparing a model system in the given ordering state,
in the vicinity of fixed density and temperature. In current models, we focus on the nematic and smectic types of ordering.
Our approach is simple enough that
the preparation of the systems of specific ordering properties can be achieved by tracking
a couple of parameters. The sensitivity of the mode, on the other hand,l is fine enough that a tiny change in
the set of parameters may produce significant changes in ordering properties.
%Performing efficient computer simulation with
%properly designed
%model mesogens at a coarse-grained level
%becomes indispensable.

To summarize, the use of the same scope to consider the ordering problem in liquid crystal and that in polymer chains enable
the addressing of the issues of interfacial problems \cite{nematicshell,dopedfibril} and the designing of appropriate models in the most efficient manner.
With the refining of the internal interactions \cite{MH3,MH6}, it is readily to extend the analysis in this paper to reproduce
quantitatively, not only the wider classes in symmetry of orderings \cite{NemaCholes,deGennes,SSRMP,SmecColmn},
but also the
important dynamic properties, to complement the laboratory observations. Such manipulations are
helpful in modeling,  to include the varieties in the liquid crystal materials \cite{NemaCholes,classHird,classGMDZC,nematicachiral},
as well as to properly reproduce the effects of interplays between the medium and the confining boundaries \cite{deGennes,spheroidconfin},
and the interfacial effects in contact with doped particles \cite{Slovakia1,Slovakia2}.

~\\
\textbf{Methods}
\vskip 2 mm

\begin{table}
\caption{Lennard-Jones interaction length parameter $L$ for the seven-monomer mesogen}
\begin{tabular}{|c||c|c|c|c|}
%{l@{\hspace{4mm}}lc@{\hspace{4mm}}c@{\hspace{4mm}}c@{\hspace{4mm}}c}
%@{\hspace{4mm}}c@{\hspace{4mm}}c@{\hspace{4mm}}c}
\hline  \hline system  &   $L_b$ for  monomers 1, 2, 6, 7   &   $L_f$ for monomer 3, 5   &    $L_c$ for monomer 4 & {$L_b$}/{$L_c$}
\\ \hline
M & $\sigma$ & $\sigma$ & $\sigma$  & 1.0000\\
\hline A & 0.876476$\sigma$ & $\sigma$ &1.14078$\sigma$   & 0.7683129\\
\hline B & 0.876332$\sigma$ & $\sigma$ & 1.14097$\sigma$  & 0.7680587\\
\hline C & 0.831258$\sigma$ & $\sigma$ & 1.20277$\sigma$ & 0.6911196\\
 \hline \hline
\end{tabular}\label{Interaction}
\end{table}

Any pair of spherical monomers,
indexed by $i$ and $j$ respectively at a distance $r_{ij}$ apart, in different molecules
or in the same molecule but not as the nearest-neighbors, is subject to the L-J
interactions,
\begin{equation}
U\left(r_{ij}\right)
=
\begin{cases}
4 \epsilon_{ij}
\left(
	\left(
	\frac{\sigma_{ij}}{r_{ij}}
		\right)^{12}
	-
	\left(
	\frac{\sigma_{ij}}{r_{ij}}
		\right)^{6}	
	\right)
+a_1 r_{ij}
+a_0
,
&
\text{if }  r_{ij} <r_{\rm c}
\\
0
,
& \text{otherwise}
\end{cases}
\label{LJpot}
\end{equation}
 where length parameter $\sigma_{ij}$ and the strength parameter $\epsilon_{ij}$ are depending on their locations in the chains and
the constants $a_1$ and $a_0$ are chosen to make the potential and the force continuous at the cutoff $r_{\rm c}$.
The soft spheres of all homogeneous and semi-homogeneous chains considered in this paper have identical L-J interactions, with
the same set of length and strength parameters, denoted by $\sigma$ and $\epsilon$. In obtaining a heterogeneous chain, we keep the
parameters for a few (identical) monomers on the chain unchanged and fine-tune the remaining monomers.
In this study, all pair of L-J interactions have the same strength parameter $\epsilon$.
For the seven-monomer  chains, we specify $\sigma$ to denote the size parameter for the inter-molecule and intra-molecule
interactions within the group of monomers, marked by 3 and 5,
(Fig.~\ref{FIG8} and Fig.~\ref{FIG9})
and those within the group of monomers 1, 2, 6, and 7 and within those monomers marked by 4, respectively, are
expressed in unit of $\sigma$, listed in Table~\ref{Interaction}.
The size parameter for the interaction between a pair of monomers from different groups is determined by the mean
$\sigma_{IJ}=\frac{1}{2}(\sigma_{II}+\sigma_{JJ})$ ($I$ and $J$ denote the groups to which monomer $i$ and
monomer $j$, respectively, belong).
Fig.~\ref{FIG8} shows the contours of inter-chain equipotential surfaces experienced by each monomer in the
semi-homogeneous seven-monomer model mesogen, for system M
plotted in Fig.~\ref{FIG5}.
Fig.~\ref{FIG9}(a)-(c) shows such contours for individual monomers for a typical heterogeneous
seven-monomer mesogen chain, which is for system B of Fig.~\ref{FIG5}.

All quantities are in units based on $\sigma$, $\epsilon$, and $m$ (mass, which are the same for all monomers.
%S
In all systems considered in this study, $\epsilon$ is also the interaction strength parameter
for any other pair of monomers, except for those nearest intramolecular neighbors.
The time unit $\tau$, specifically, is defined by $\tau=\sqrt{\frac{m\sigma}{\epsilon}}$.
We use the superscript ``$^*$'' to label the ``reduced'' quantities,
which are quantities in units of $\sigma$, $\epsilon$ and $m$,
or their composites.
The reduced temperature $T^*$, the reduced density $\rho^*$, and the reduced pressure $P^*$,
for example, are defined by
$T^*=k_{\rm B}T/\epsilon$,
$\rho^* =\rho \sigma^3$,
and $P^*=P\epsilon^{-1}\sigma^{3}$, respectively.
Note that, the reduced quantities of systems A, B, and C, are defined in terms of the
parameters shared by the two monomers next to the center of a chain (see Table~\ref{Interaction}).
In some figures, specifically,
in Fig.~\ref{FIG5}, the superscript $^*$ of some reduced quantities are omitted, for the clearness of the plots.

We use RATTLE algorithm \cite{RATTLE} to maintain each bond between any two nearest neighboring monomers along a
chain, except for those connections to the end monomers in the seven-monomer chains, strictly
to have a constant length $b=0.357 \sigma$.
%The chain length $n=10$ is only one-tenth of %systems of chains, with the same interaction
%parameters, with ten times number of sites ($n=100$) in each chain,
For those homogeneous chains of lengths $n=100$ considered in
the study of polymer aggregation \cite{MH3} where the system has the same interaction
parameters as those in Figs.~\ref{FIG1} and~\ref{FIG2},
 %have the %been} shown that the
the system has a persistent length $l_p$ approximately fifteen times of the bond length $b$. %of the chains,
$l_p$ is calculated by finding the decaying distance of the
direction cosine
$
\left \langle
	\hat{b}_i \cdot \hat{b}_{j}
	\right \rangle
$
between the bond directions $\hat{b}_i$ and $\hat{b}_j$,
over any pair of monomers indexed sequentially by $i$ and $j$ along individual chains,
in fitting to
$\left \langle
	\hat{b}_i \cdot \hat{b}_{j}
	\right \rangle
=
e^{-\frac{l}{l_p}}
$ as a function of
the (curved) distance $l=j-i$ in between.
The tail monomers on the two ends of a seven-monomer mesogen are connected
by a spring potential $V(r_{ij})= \frac{1}{2} k_{\rm 0} (r_{ij}-b)^2$, where $k_0=1.5552 \times 10^3 \epsilon$.
In simulation, the cutoff $r_c$ is $2.3\sigma$ for all cases. The integration time step, in solving the equations of motion,
is $1.0 \times 10^{-3} \tau$ for the two systems of homogeneous chains, described in Fig.~\ref{FIG1} (each chain of length $n=10$)
and Fig.~\ref{FIG2} ($n=5$), respectively, and
is  $1.0 \times 10^{-4} \tau$ for all systems ($n=7$) considered in Fig.~\ref{FIG5}.

In applying an external field, a potential $u_F$ is assigned for each chain, which is a function of the inner product
between the direction of external field $\hat{H}$ and the induced dipole \cite{deGennes}
$\vec{M}=\chi_{\bot} (\vec{H}- \hat{n}\vec{H}\cdot \hat{n})+\chi_{\|} \vec{H}$, with $\chi_{\|}<0$ and $\chi_{\bot}<0$,
at the core segment of the chain, where $\hat{n}$ is determined by
the relative position of two neighboring monomers $I$ and $J$ at the center of the chain,
$\hat{n}=\frac{\vec{r}_I-\vec{r}_J}{|\vec{r}_I-\vec{r}_J|}$.
The gradients of the field-driven potential $-\vec{H}\cdot \vec{M}$ with respect to the positions $\vec{r}_I$ and  $\vec{r}_J$,
respectively, produce a torque to
turn the core segment of the chain toward the direction of the field. In simulation, we use an auxiliary field in $z$-direction,
\begin{equation}
 u_F(\vec{r}_I,\vec{r}_J)=g_0- g_2(\hat{z}\cdot \hat{n})^2,
\label{field}\end{equation}
with $g_2=15$ (in reduced unit)
in the turning the field on shown in the plane M in Fig.~\ref{FIG5}, over a time duration of $104 \tau$. In testing the stability,
a stronger field, $g_0=150 $
is applied in each case. Since the purpose of applying the field in the test is simply to provide an appropriate initial
configuration for a fast relaxation, the time duration of the field is over a short time span which is
sufficient to arrange the orientations of the chains to follow the direction of the field.

Starting with an initial configuration, without any knowledge on the status of ordering in the equilibrium state after relaxation,
an isotropic isobaric process is carried out by
using the extended system method \cite{AT87}, to balance the
instantaneous pressure $P^*$ by imposing an externally effective mass, in such a way that the deviation from a target value $P_0$ exerts
an acceleration to the volume of the system to compensate that deviation.
In thermodynamic equilibrium, the setup would generate the realization of $NPH$ \
($N$: particle number, $P$: pressure, and $H$: enthalpy) ensemble.
In a quenching process, we take the heat out of the system
by rescaling the velocities of the monomers on the chains, in accord with a decreasing target temperature $T_0$.
In the system composed of $N=800$ homogeneous chains, each with $n=5$ monomers (Figs.~\ref{FIG3}),
we take the heat-removing actions in intervals of 1250 steps ($1.25 \tau$), with the target temperature $T_0$ at each
heat-removing step changing to a ratio 0.9999 of the value at the previous action.
In between two heat-removing steps, the system is in isobaric evolving ($NPH$).
 For the quenching process carried out in system M of Fig.~\ref{FIG5}, the ratio is 0.99995 and the time interval of heat-removing
is 1000 steps ($0.1 \tau$). In contrast to the case in Fig.~\ref{FIG3}, the system of chains of seven-monomer ($n=7$), with
spring-connected tails, is in non-equilibrium and spatially highly non-uniform over the process. While, before the pressurizing step,
the target pressure is set at $P_0 = 1.0 \times 10^{-4}$ and $P_0 = 1.0$, respectively,
before and after turning off the field, the system stays within the unstable coexistence regime with its true pressure in negative value.
Such a situation is realized as a consequence of on-going re-arrangement process in the partially filled space and it
allows the molecules to reorient efficiently to the direction of the imposed field.

\vskip 4 mm
~\\
\textbf{Acknowledgements}
	This work was supported by the grants MOST 104-2811-M-001 -121 (JB),
	MOST 104-2923-M-001 -004, MOST 106-2112-M-001-027, MOST 107-2112-M-259-006
and MOST 107-2911-I-004-503, MOST 108-2112-M-259 -008 and MOST 108-2811-M-259-505 (SCW) of Taiwan,
  also the grants APVV SK-TW 2017-0012 and APVV 15-0453 of Slovakia.

  \vskip 3 mm
  ~\\
\textbf{Author Contributions}
W.J.M., S.C.W., J.B., N.T., P.K. and C.K.H. discussed and prepared the primary model. W.J.M. designed the working model.
W.J.M. and LW.H performed the simulation. W.J.M,. L.W.H,. S.C.W. and C.K.H. carried out data analysis. W.J.M. and C.K.H.
wrote the manuscript taking into account comments from S.C.W., J.B. N.T., M.T., P. K. and K.S.,

\vskip 3 mm
~\\
\textbf{Additional Information}
Competing Interest. The authors declare that they have no competing interest.

\vskip 10 mm

%\vskip 6 mm

\newpage

%% Fig. 1
\begin{figure}
\begin{center}
%%% (here is Fig. 1)
%%% \includegraphics[width=0.45\textwidth]{FIG1A.eps}
%%% \includegraphics[width=0.45\textwidth]{FIG1B.eps}
\end{center}
\caption{(a) Snapshot of a system with $N=400$ homogeneous chains, each chain having $n=10$ monomers. The system is at
$T^*\equiv \frac{k_{\rm B}T}{\epsilon}=1.067941$
in a $L \times L \times L$ cubic box with  $L=18.9997 \sigma$. The main body of the configuration consists
of a layer of nearly-straight parallel
chains in a crystalline structure. Under periodic boundary conditions, each chain in the layer
is plotted in two separate parts, next to two opposite faces of the cubic simulation box. There are two smaller ordered
clusters attaching to the main layer, with their orientations in parallel to that plane.
Those clusters occupy only, approximately, one-quarter of the total volume of the box, leaving the rest space empty.
(b) Projection of (a) in two dimensions, viewed in the
direction of the backbones of the chains. Each monomer of the chain is plotted as a filled circle.
Each chain in the main layer has two separate clusters of overlapped images,
each of which are contributed by those monomers belong to one of the two parts plotted in (a). Those monomers
in each cluster of images overlap with one another very well since the backbone is nearly straight.
Each blue line marks the projection of a inter-molecular connection between one of the two monomers at the center of the ten monomers
of a chain and its, inter-chain, nearest neighbor monomers (not necessarily at the center of the other molecule).
We plot only those connections for the part of the chains
plotted next to the left face of the box in (a). We
can see that those blue lines combined to form a six-fold triangular lattice. The projected images of those monomers next to
the right face of the box in (a) are however, not on the lattice.
The results indicate the directions of the backbones tilt from the normal direction of the plane of the layer.
The orientation ordering parameters obtained by diagonalizing Eq.~(\ref{order}) for the molecules,
are $S=0.817$ and $\eta=0.0465$, respectively. Those for the bonds (Eq.~(\ref{orderbond})) are
$S^{\rm bond}=0.813$ and $\eta^{\rm bond}=0.0462$, respectively.
} \label{FIG1}
\end{figure}

\newpage

%% Figure 2
\begin{figure}
%% \begin{figure}[tb]
\begin{center}
%%% (here is Figure 2)
%%% \includegraphics[width=0.45\textwidth]{FIG2A.eps}%g68335000.eps}
%%% \includegraphics[width=0.45\textwidth]{FIG2B.eps}%g68335000.eps}
\end{center}
\caption{Snapshot of a system with $N=800$ homogeneous chains; each chain has $n=5$ monomers.
Each bond connecting a pair of the nearest neighbor monomers along a chain is plotted as a straight line.
Similar to Fig.~\ref{FIG1}, the intra-chain nearest neighbor bonds of
some of the molecules are plotted dividedly into two separately connected
pieces in two opposite faces of the box, so as to arrange all monomers of a chain inside
the cubic box, under periodic boundary conditions (see Fig.~\ref{FIG1}).
In the plots, we show how the radial distribution functions (see Fig.~\ref{FIG4}) are calculated from the
projections of the chain-molecules in a configuration, (a) on the plane
in perpendicular and (b) on the line in parallel, respectively,
to the direction of any reference molecule (marked by a thick segment of green line in each plot).
In (a), those molecules with their centers in the cylindrical
shell centered at the line passing through the center in the direction vector of the reference molecule, of
radius $r_{\bot}$, height $h$ %$l_s$
and thickness $\Delta r_{\bot}$, are projected into the circular ring of radius $r_{\bot}$
and of width $\Delta r_{\bot}$ on the normal plane.
In (b), those within the circular slab of radius $r_s$, thickness $\Delta r_{\|}$ and distance $r_{\|}$
away from the normal plane are projected into the interval of width $\Delta r_{\|}$ at the distance $r_{\|}$
away from the center and in the direction of the reference chain.  The configuration is at
instantaneous temperature $T^*\equiv \frac{k_{\rm B}T}{\epsilon}=1.8887$
with a cubic volume of size $12.6651 \sigma$ on each edge, over an isobaric process of controlling pressure
$P^* \equiv P \epsilon^{-1}\sigma^3 =  0.1$ (marked by diamond in recording the data of quenching
in Fig.~\ref{FIG3}).
We calculate the orientation ordering parameters by diagonalizing Eq.~(\ref{order}) for molecules,
to obtain $S=0.975$ and $\eta=6.94 \times 10^{-4}$. Those for bonds (Eq.~(\ref{orderbond})) are $S^{\rm bond}=0.969$ and
$\eta^{\rm bond}=6.34 \times 10^{-4}‬$ for bonds.
} \label{FIG2}
\end{figure}

\newpage

%% Figure 3.
\begin{figure}
%% \begin{figure}[tb]
\begin{center}
%%% (Here is Figure 3)
%%% \includegraphics[width=0.95\textwidth]{FIG3.eps}
\end{center}
\caption{Data of instantaneous temperature $T^*$, instantaneous pressure $P^*$, density $\rho^*$ and the parameter
$S$ for the quenching process of the system of  Fig.~\ref{FIG2}: (a) $S$ vs. $\rho^*$, (b) $T^*$ vs. $S$, (c) $T^*$ vs. $\rho^*$,
(d) $T^*$ vs. $P^*$, (e) $P^*$ vs. $S$ and (f) $P^*$ vs. $\rho^*$.
The last time spots of the simulation are marked by an open circle (with $T^*=1.42$ and $\rho^*= 0.404$ in (c)).
The track of the process is guided by the red curve in (c), (d), and (f).  The quantities are
obtained by averaging over intervals of 0.2 $\tau$= 200 steps.
The data points (filled black circles) are plotted every $10 \tau$=10000 steps.
The line from a triangle (with $T^*=4.58$ and $\rho^*=0.406$ in (c)) to a square (with $T^*=4.16$ and $\rho^*=0.243$ in (c))
represents the crossover from the vapor to the liquid state (at $T^* \approx 4.369$ in (c))
with abrupt increasing in the amplitude of the fluctuation in $P^*$ (in plots (d), and (f)) and $\rho^*$;
 (d) and (f) indicate that the system is maintained in
thermodynamic equilibrium during the process, where the time-averaged
instantaneous pressure (red line) is always maintained at the target pressure ($P_0$  =  0.1 in unit of $\epsilon \sigma^{-3}$,
marked by solid blue lines in (d), (e) and (f)), on both the vapor side and the liquid side.
Another occasion of abrupt change in the value of $\rho^*$ is accompanied
by the change in the value of $S$, in between the points marked by the '$\times$'  (with $T^*=2.15$ and $\rho^*=0.354$)
and the diamond (with $T^*=1.91$ and $\rho^*= 0.394$), at $T^* \approx 2.032$ in plot (c).
It signals the crossover from liquid to crystal.  The time-averaged
instantaneous pressure (red lines in plots (d) and (f)) %(red lines in the right plot and in the bottom plot),
is apparently smaller than the
target value ($P_0$ = 0.1, marked by solid blue lines) at this crossover and is even negative ($P_0$ = 0.0
is marked by dashed blue lines in (d), (e) and (f)) on entering the crystal
regime. The system is, therefore, in a non-equilibrium situation over the process in entering this regime, under such a fast quenching.
} \label{FIG3}
%there is a large change at the crossover from liquid to solid %(top plot and left plot),
%at $T^* \approx 2.032$}.
%% We mark two time spots, each before or after this change, by a '$\times$' symbol and a diamond (in red color), respectively.
%% the radial distributions at which, among those at the other labelled  spots, are analyzed in Fig.~\ref{FIG4}.
\end{figure}

%% \widetilde{}

\newpage
%% Figure 4.
\begin{figure}
\begin{center}
%%% (Here is Figure 4.)
%%% \includegraphics[width=0.97\textwidth]{FIG4.eps}
\end{center}
\caption{Radial distribution functions (a) $g_{\|}(r_{\|})$ (Eq.~(\ref{gperpen}) for the projection on the plane perpendicular to the direction of
the reference molecule and Fig.~\ref{FIG2} (a)),
(b) $g_{\bot}(r_{\bot})$ (Eq.~(\ref{gpara}) for the projection on the line passing the center of mass, along the direction of the reference molecule
 and Fig.~\ref{FIG2} (b)), (c) $g_{\rm cos}(r_{\rm site})$ for the averaged direction cosine between
the bonds for a pair of sites at a distance $r_{\rm site}$ from each other,
computed for the system of 800  five-site homogeneous chains, at the five time spots
specified in
the plots of time evolution over the quenching process described by
 Figs.~\ref{FIG3} and \ref{FIG4}. The lines in gray, red, green, blue and black
colors are data for those time spots in Figs.~\ref{FIG3} and \ref{FIG4}, marked by triangle, square,
``$\times$'', diamond and
open circle, respectively. The plots show that the system start to develop
longer range orders, on entering the crystal %solid
regime (blue and black lines in plots (a)-(c)). While the translation order
continues to increase,
in lowering the temperature further down in the crystal
regime
(black lines in plot (a) and (b)), the orientation order
immediately reaches the saturated level once  entering the regime (blue line in plot (c)). At the end of the simulation
(black lines), the system possesses a structure of
stacked planes of triangular lattices. This is evidenced by the match-up between the locations of the
major peaks in $g_{\bot}(r_{\bot})$
and those produced by an ideal planar
triangular lattice (marked by the dashed vertical lines in plot (b)). The presence of the broad peak in
between $r=0$ the location of the main (highest) peak
indicates the orientation of the molecules are tilt from the normal of the stacked planes. We put the equal-spacing dashed vertical lines
in plot (c), to mark the
locations of the neighbor sites along a mechanically fully relaxed five-site chain, for reference.
} \label{FIG4}
\end{figure}

\newpage
%%% Figure 5.
\begin{figure}
%% \begin{figure}[tb]
\begin{center}
%%% (Here is Figure 5.)
%%% \includegraphics[width=0.71\textwidth]{FIG5.eps}
\end{center}
\caption{Schematic charts to show the transfer from System M with $N=2000$ seven-monomer mesogens to Systems A, B, and C,
with smectic, nematic and isotropic fluid states, respectively, marked by blue ``+'' symbols.
The data of instantaneous reduced temperature $T^*$,  versus reduced density $\rho^*$ for a quenching and squeezing process, carried out in
the coexistence region of System M, is shown (black dots) every 1.0 $\tau$ on the plane marked by
``M''.
For the simplicity of notation, we omit in the plots the superscript ``$^*$'' in denoting those variables which are in reduced units (see Methods).
The gray dashed line on M is the trajectory of averaging the data for the five-monomer ($n=5$) mesogens of Fig.~\ref{FIG3} and are used as
a reference to estimate the boundary of the coexistence region for the system M. The red thick line on M is the trajectory of time averages over
intervals of 10 %50
consecutive data points (a time span of $10 \tau$ We present the
side views (projections on x-y planes) for a sequence of transient non-equilibrium snapshots over the quenching process for system M
(labeled by $M_1$, $M_2$, $M_3$, $M_{\rm I}$ and $M_{\rm II}$, with their locations on plane M marked by open circles).
The orientation-ordered configurations are
obtained by applying a pulse of applied auxiliary field in z-direction, (the on-off marked by
``field on'' and ``field off''), followed by imposing
pressure to $P_0=20.0$ (at the spot marked by ``pressurizing on'') at the controlled temperature $T^* \approx T_0 $=3.33675,
to allow the space to be fully filled by the molecules.
It is then followed by tuning the L-J interaction parameters,
to reach in, via the pathways I or II, the stable smectic, nematic and isotropic fluid configurations at, respectively,
the points %(marked by blue "+")
$(\rho_{\rm I}, T_0)_{\rm A}$, %(on plane "A"),
$(\rho_{\rm II}, T_0)_{\rm B}$, %(on plane "B")
and $(\rho_{\rm I}, T_0)_{\rm C}$ (those points, marked by ``+'' on the corresponding planes),
%(on plane "C"),
where $\rho_{\rm I}$ =0.313933 and $\rho_{\rm II}$ =0.348452.
%and also those of the relaxed stable or metastable configurations, for systems A, B and C (those points, marked by "+" on the corresponding planes),
%which are reached from the initial configurations $M_{\rm I}$ or $M_{\rm II}$ on M,
%via the pathways I or II, crossing the space of interaction parameters.
The details are  presented in the text.
}\label{FIG5}
\end{figure}

\newpage
%% Figure 6
\begin{figure}
%% \begin{figure}[tb]
\begin{center}
%%% (Here is Figure 6.)
%%% \includegraphics[width=0.95\textwidth]{FIG6.eps} %g69655000.eps}
\end{center}
\caption{Radial distribution functions (a)$g_{\|}(r_{\|})$,
(b) $g_{\bot}(r_{\bot})$ and (c) $g_{\rm cos}(r_{\rm site})$, for the configurations
at  $(\rho_{\rm I},T_0)_{\rm C} $ (gray lines),
$(\rho_{\rm I},T_0)_{\rm A} $  (red lines) and
 $(\rho_{\rm II},T_0)_{\rm B} $ (green lines).
In all cases, except for the isotropic configuration $(\rho_{\rm I},T_0)_{\rm C}$,
the functions $g_{\rm cos}(r_{\rm site})$
have high orientation correlations, extended over the whole ranges of $r_{\rm site}$ shown in
(c) (the locations of the neighbor sites along a fully relaxed chain marked by vertical dashed lines).
For the configurations the isotropic
and nematic
configurations
at  $(\rho_{\rm I},T_0)_{\rm C}$ and $(\rho_{\rm II},T_0)_{\rm B}$, respectively, % in Fig.~\ref{FIG5}),
their functions $g_{\|}(r_{\|})$ are flat
indicating no layered-structures.
For comparison, we also put the data
(black lines) for a smectic configuration (a state marked by diamonds in Figs.~\ref{FIG3})
in the five-monomer %site
system (Fig.~\ref{FIG2}), which have been shown
(blue lines)
in Fig.~\ref{FIG4}.
For the smectic configuration at $(\rho_{\rm I},T_0)_{\rm A}$,
the apparent smaller
amplitudes in the oscillations among peaks and valleys
in its $g_{\|}(r_{\|})$ than those for
the smectic five-monomer-chain configuration %(R)
can be understood in terms of the more extended range in the layered structures for
the latter than it is for the former. %$(\rho_{\rm I},T_0)_{\rm A}$.
The match-up between the locations of the peaks in $g_{\bot}(r_{\bot})$ and
the those determined by a perfect planar triangular lattice (red dashed vertical lines, based on the location of the main peak)
fails beyond the distance of the fifth peak away from $r_{\bot}=0$, % the origin of $r_{\bot}$,
in contrast to the agreement
between the lattice (black dashed vertical lines) and the function in the case of five-site configuration.
For the case of nematic configuration prepared via pathway I (Fig.~\ref{FIG5}),
the curves are virtually identical with
those (blue dots) for a configuration %relaxed
obtained via a different pathway
$
(\rho_{\rm I},T_0)_{\rm C} \rightarrow (\rho_{\rm I},T_0)_{\rm B} \rightarrow (\rho_{\rm II},T_0)_{\rm B}
$
reaching at a point with a slightly different density}\label{FIG6}
\end{figure}

\newpage
%% Figure 7
\begin{figure}
%% \begin{figure}[tb]
\begin{center}
%%% (Here is Figure 7.)
%%% \includegraphics[width=0.255\textwidth]{FIG7a.eps}
%%% \includegraphics[width=0.255\textwidth]{FIG7b.eps}
%%% \includegraphics[width=0.255\textwidth]{FIG7c.eps}
\end{center}
\caption{Side views (projections on x-y planes) of the configurations under relaxation after applying a
pulsed field (see Methods),
for system A in Fig.~\ref{FIG5}. Before applying the field, the system was obtained by
starting from a melted isotropic configuration
of system B, at the point
$(\rho_{\rm I}, T_0)_{\rm B}$ of system B.
It then underwent a process changing the interaction parameters over pathway I reversely
(Fig.~\ref{FIG5}), to reach the point
 $(\rho_{\rm II}, T_0)_{\rm A}$ on plane A (Fig.~\ref{FIG5}), followed by a process reducing its density to reach $\rho^*=0.348356$
in the vicinity of the point $(\rho_{\rm I}, T_0)_{\rm B}$, where $\rho_{\rm I}$  equals 0.313933 (Fig.~\ref{FIG5}).
The plots show the snapshots (a) in the beginning of the relaxation; (b) at a time
after $83 \tau$; (c) at a time
after $205 \tau$.
We see that the system develops lamellar structures, that merits the smectic state. The system is still on
relaxation and the presence of many
defects at molecular scales keep the configuration not identical to the one at $(\rho_{\rm I}, T_0)_{\rm B}$ in Fig.~\ref{FIG5}.
}\label{FIG7}
\end{figure}
\newpage

%% Figure 8
\begin{figure}
\begin{center}
%%% (Here is Figure 8.)
%%% \includegraphics[width=0.55\textwidth]{FIG8.eps}
\end{center}
\caption{Contour of the inter-chain equi-potential surfaces produced by a seven-monomer model mesogen of system M mentioned
in Fig. \ref{FIG5} and Table \ref{Interaction}. In the seven-monomer model mesogen,
two extra monomers  are added via two  springs to two sides of a core composed of rigidly-bonded five-monomer homogeneous chain,
considered in Figs.~\ref{FIG2} to Figs.~\ref{FIG4}. All seven monomers are identical in their intermolecular interactions.
The contours are the cross-sections of the equipotential surfaces on the $x$-$y$ plane.
The potential is for one monomer of any chain, produced by another chain which is fully relaxed on
the $y$-axis, with the center monomer placed at the origin.
It is the sum of seven identical L-J pair
potentials, Eq.~(\ref{LJpot}), with length parameter $\epsilon$ and strength parameter $\sigma$,
between the given monomer of the first chain and the seven monomers of the second chain, placed on the positions (marked by dots) on the plot.
The values on the plot is in unit of $\sigma$.
Two nearest neighboring monomers are at a distance of $b$=0.357$\sigma$ on the chain.
The model molecule has an effective aspect ratio $0.357\times 6+1.0=3.142$. All the seven-monomer model mesogens considered in this
article are obtained by tuning such a potential according to the parameter of Table \ref{Interaction}.
} \label{FIG8}
\end{figure}

\newpage

%% Figure 9
\begin{figure}
%% \begin{figure}[tb]
\begin{center}
%%%(Here is Figure 9.)
%%% \includegraphics[width=0.465\textwidth]{FIG9a.eps} %g69655000.eps}
%%% \includegraphics[width=0.465\textwidth]{FIG9b.eps} %g69655000.eps}
%%% \includegraphics[width=0.465\textwidth]{FIG9c.eps} %g69655000.eps}
\end{center}
\caption{
Contours of the inter-chain equipotential surfaces experienced by each monomer in a typical seven-monomer mesogen,
of system B mentioned in Fig.~\ref{FIG5} and Table \ref{Interaction}.
The contours are the cross-sections of the equipotential surfaces on the $x$-$y$ plane.
The potentials are produced by a fully relaxed chain placed along the $y$-axis, with the center monomer at
the origin similar to Fig.~\ref{FIG8}.
%%  and is the sum of seven monomer-monomer interaction pair potentials, which are heterogeneously and symmetrically
%%  arranged on the two sides of the center monomer.
The potential surfaces are obtained by assigning the L-J potentials (Eq.~(\ref{LJpot})).
We assign the same L-J potential for the monomer-monomer intermolecular as well as non-nearest neighboring intramolecular
pair interaction among the group of monomers marked by 3 and 5. Its
length parameter $\sigma$ and strength parameter $\epsilon$  are taken as the units for the rest potentials.
We assign the same potential for the monomer-monomer intermolecular as well as non-nearest neighboring intramolecular interactions
of any pair among the group of monomers marked by 1, 2, 6, or 7,
and its length parameter is 0.876332$\sigma$. The L-J pair potentials for those center monomers, marked by 4, have the length parameter, 1.14097$\sigma$.
The cross group pair interaction potentials are determined by assigning, respectively, the length parameter
as the algebraic mean and the strength parameter as the geometric mean, of the corresponding values within the two groups.
The monomers marked by 1, 2, 6, and 7 experience
the inter-molecular interaction produced by another molecule, with the contours of projection of potential surface
on x-y plane plotted in (a).
Plots (b) and (c) are for monomers marked by 3 and 5 (for (b)) and 4 (for (c)), respectively.
For system A and system C in Fig.~\ref{FIG5}, the strength parameters are again $\epsilon$, the same for all pairs of monomers.
Together with the length parameter for the group of monomer 3 and 5; all quantities in units of $\epsilon$ and $\sigma$
are listed in Table~\ref{Interaction}.}
\label{FIG9}
\end{figure}

%% Fig. 1
\begin{figure}
\begin{center}
(here is Fig. 1)
\includegraphics[width=0.45\textwidth]{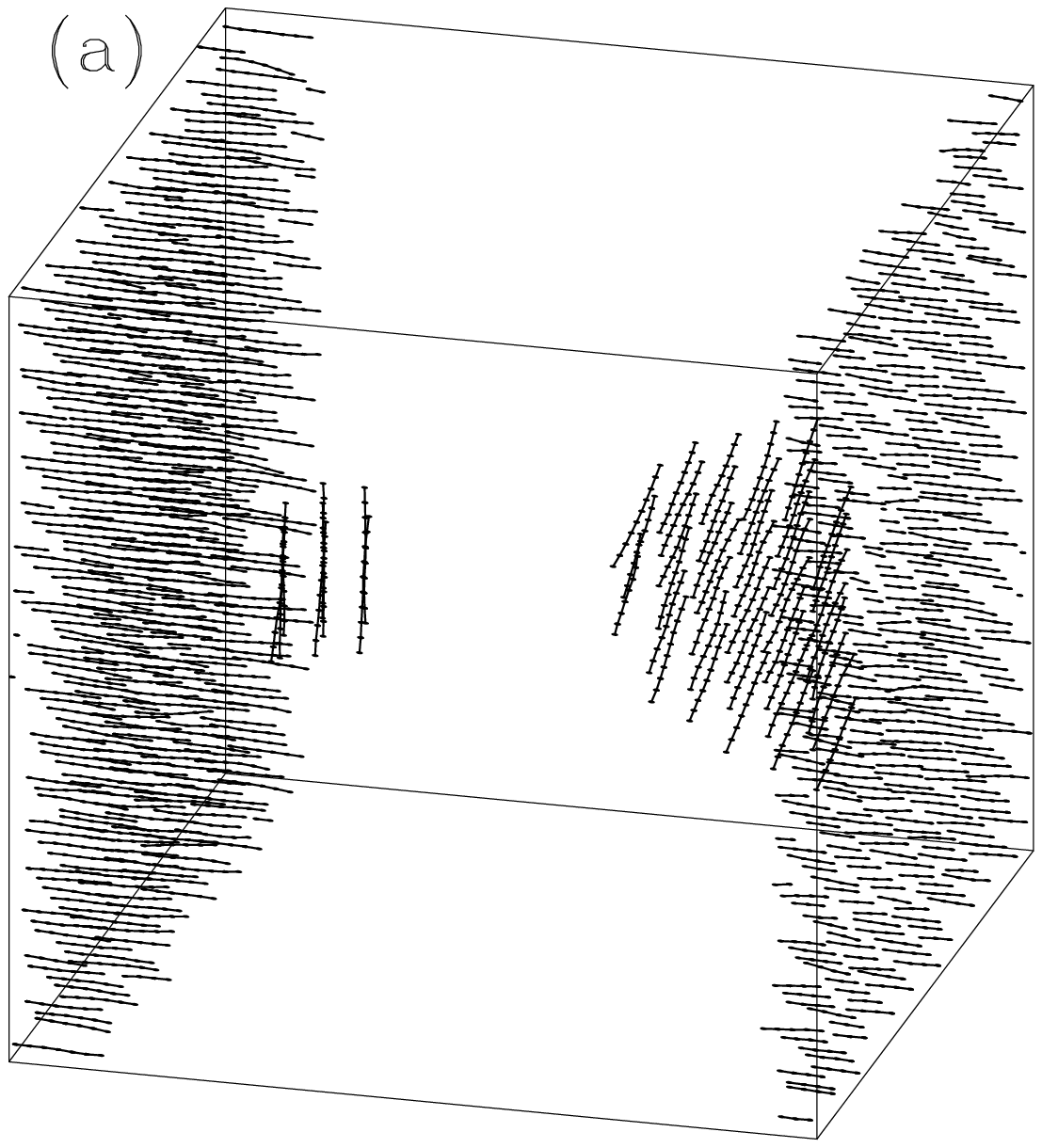}
\includegraphics[width=0.45\textwidth]{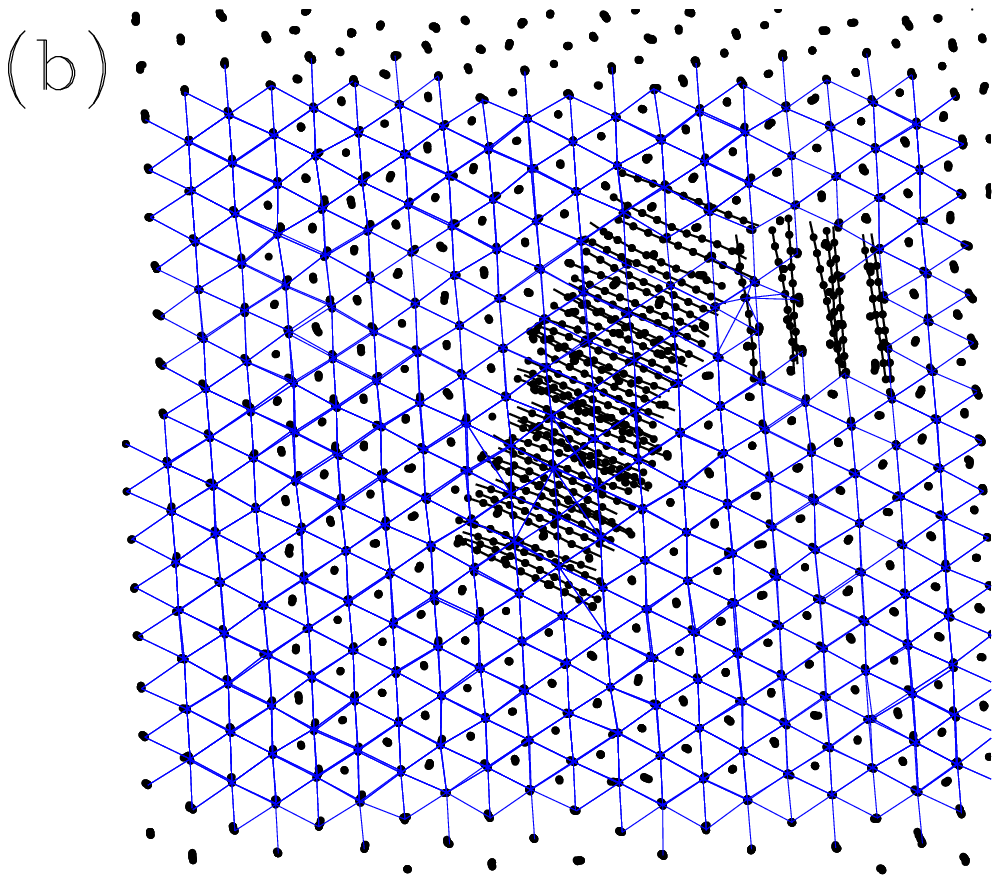}
\end{center}
\end{figure}

%% Figure 2
\begin{figure}
%% \begin{figure}[tb]
\begin{center}
 (here is Figure 2)
 \includegraphics[width=0.45\textwidth]{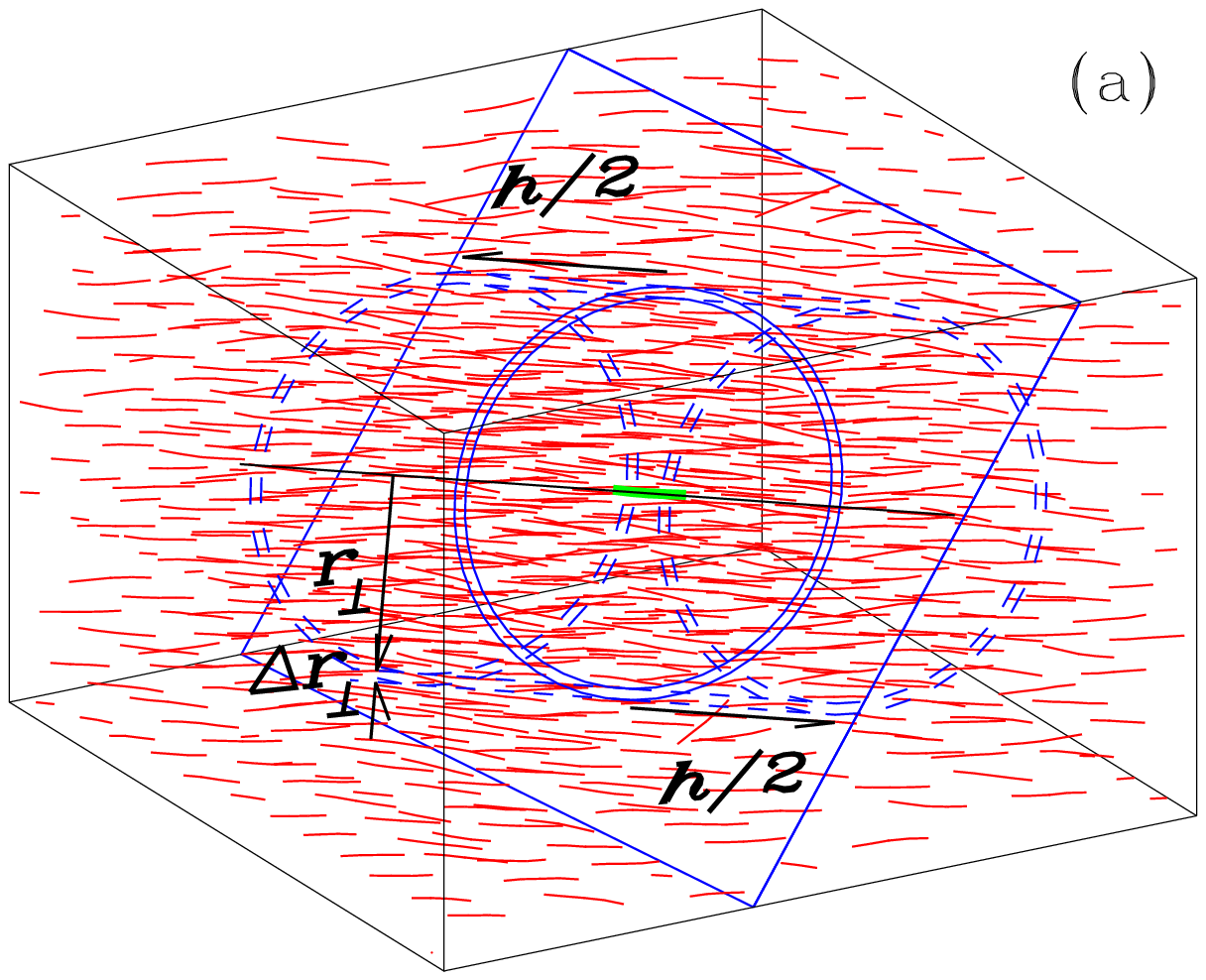}%g68335000.eps}
\includegraphics[width=0.45\textwidth]{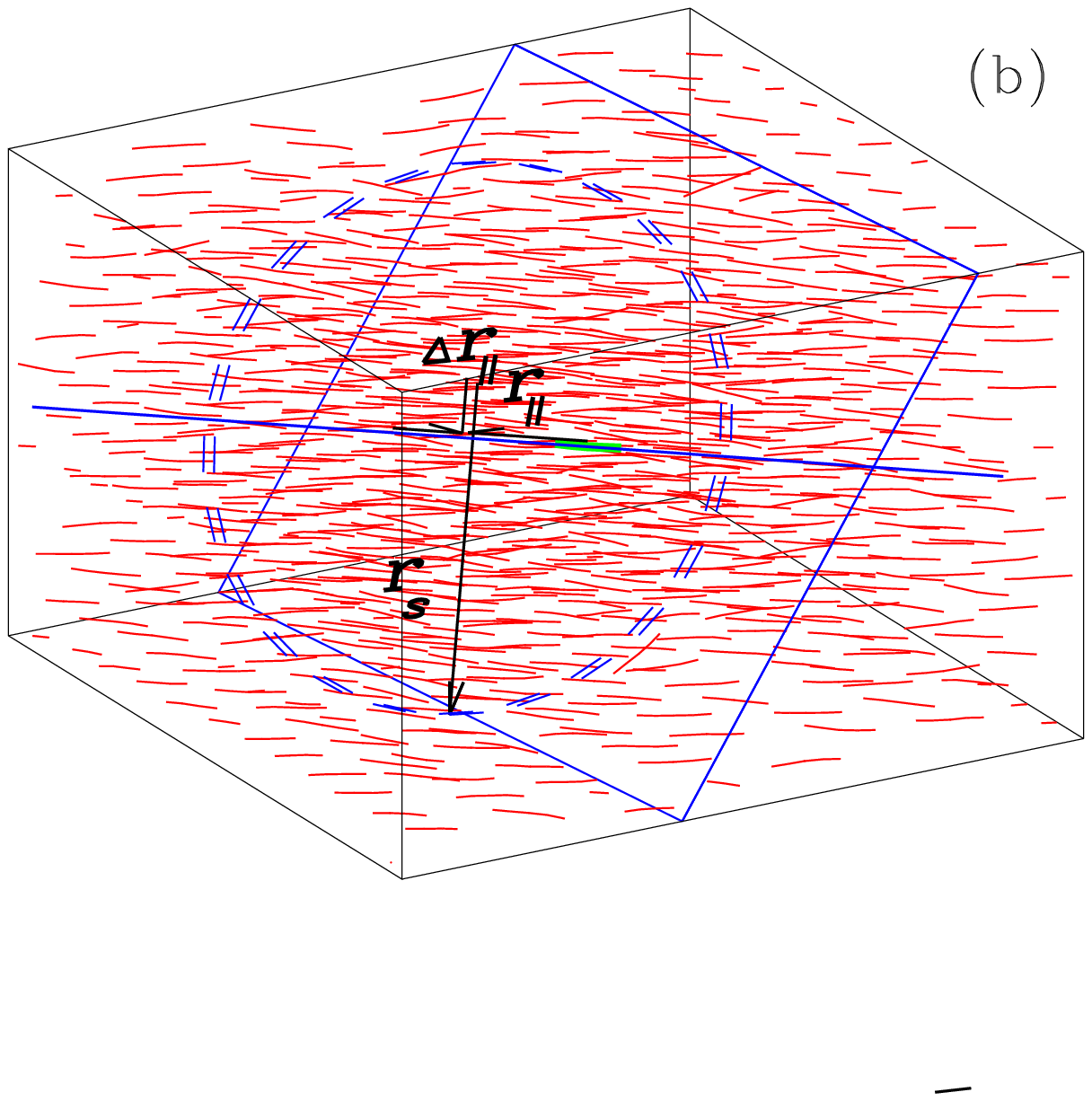}%g68335000.eps}
\end{center}
\end{figure}

%% Figure 3.
\begin{figure}
%% \begin{figure}[tb]
\begin{center}
(Here is Figure 3)
\includegraphics[width=0.95\textwidth]{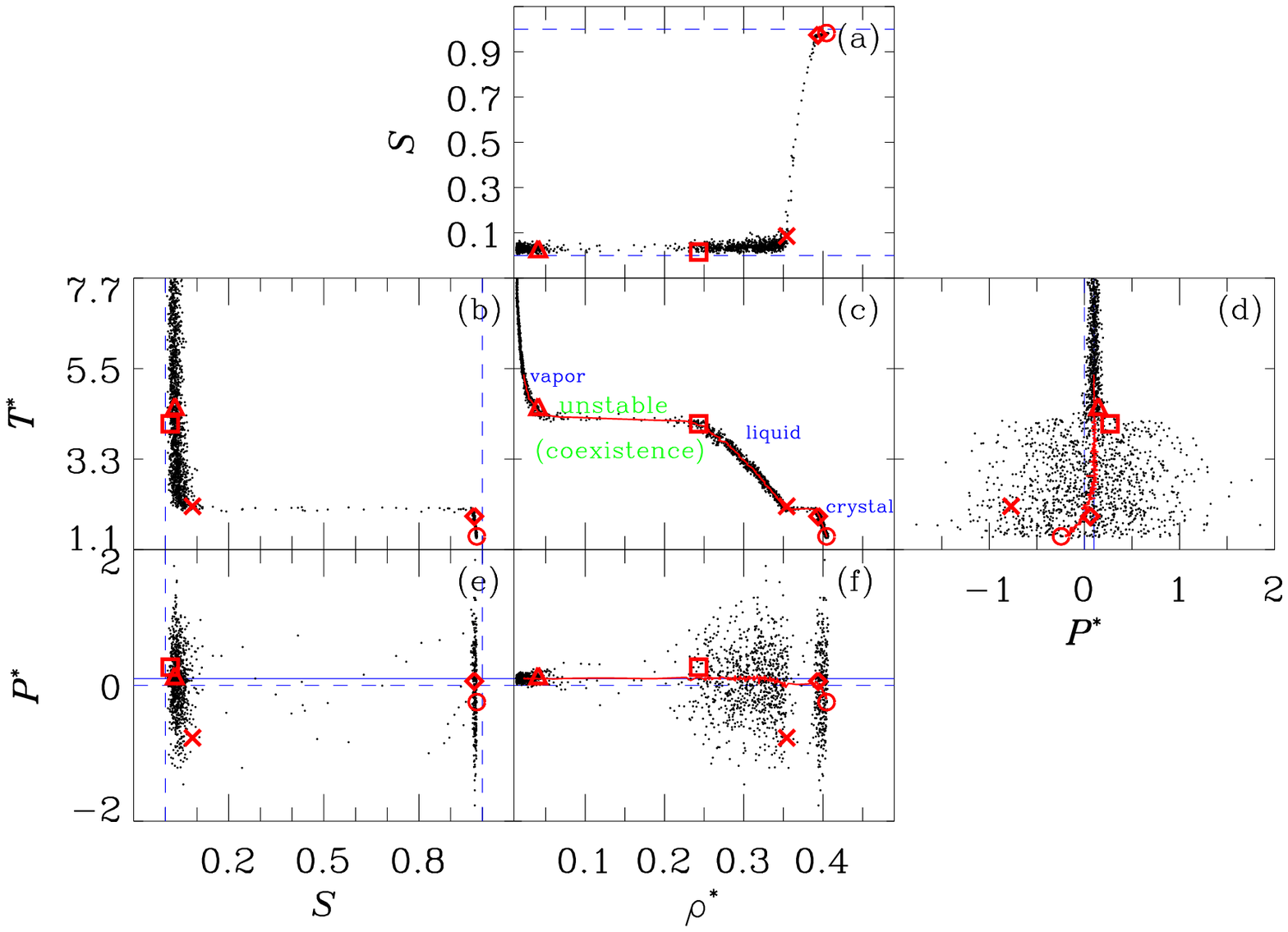}
\end{center}
\end{figure}

%% Figure 4.
\begin{figure}
\begin{center}
(Here is Figure 4.)
 \includegraphics[width=0.97\textwidth]{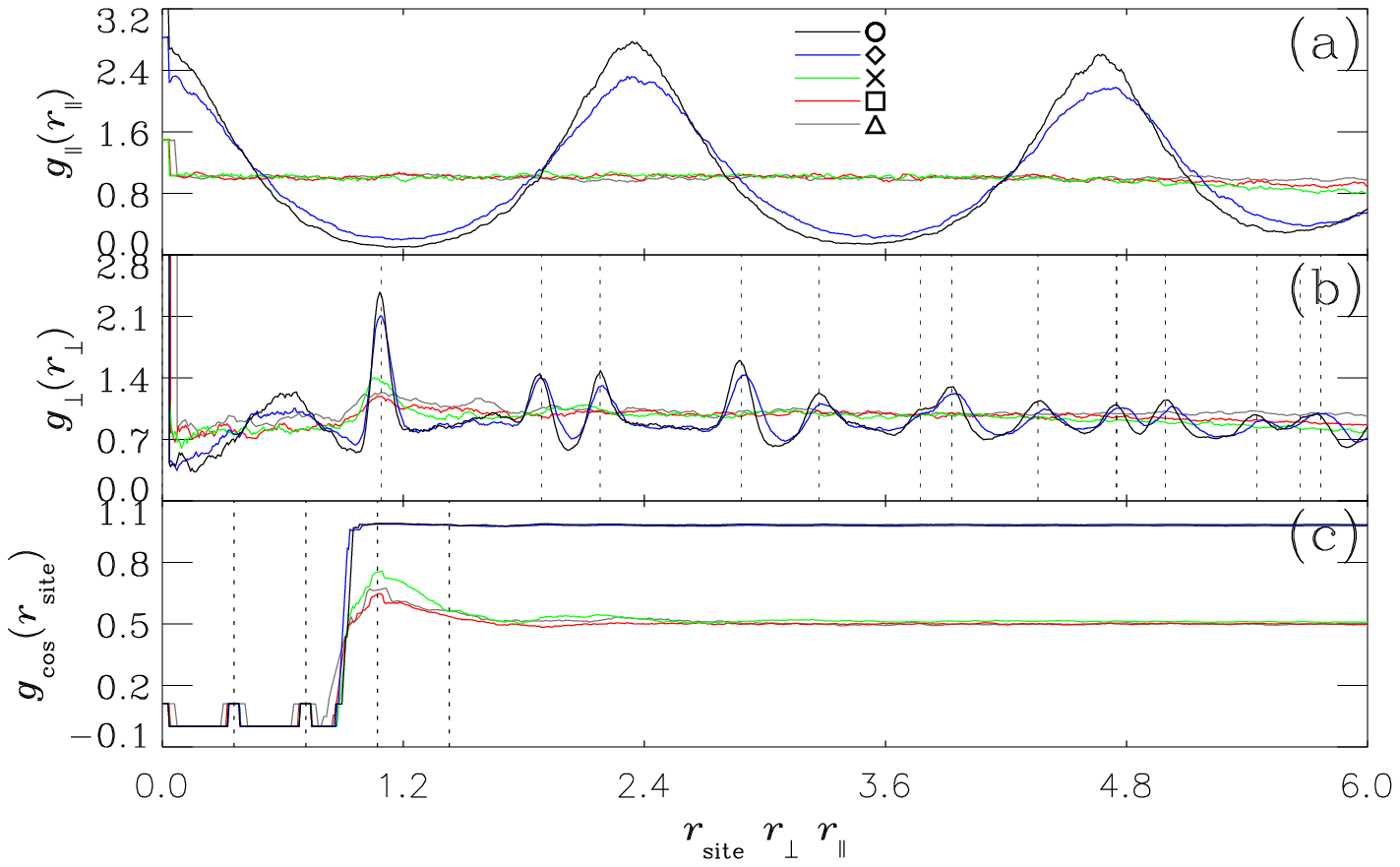}
\end{center}
\end{figure}

%%% Figure 5.
\begin{figure}
%% \begin{figure}[tb]
\begin{center}
 (Here is Figure 5.)
 \includegraphics[width=0.71\textwidth]{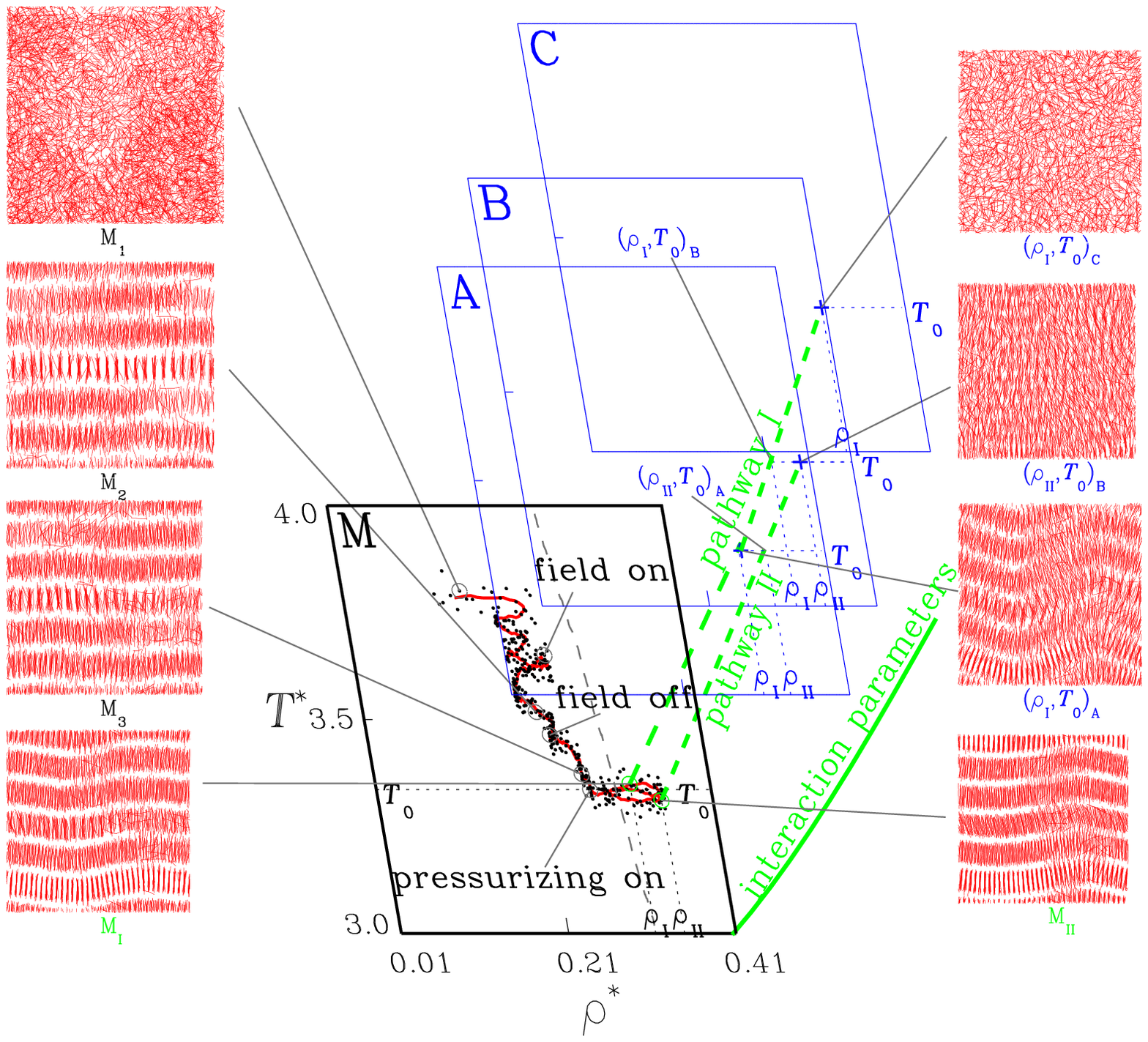}
\end{center}
\end{figure}

%% Figure 6
\begin{figure}
%% \begin{figure}[tb]
\begin{center}
 (Here is Figure 6.)
\includegraphics[width=0.95\textwidth]{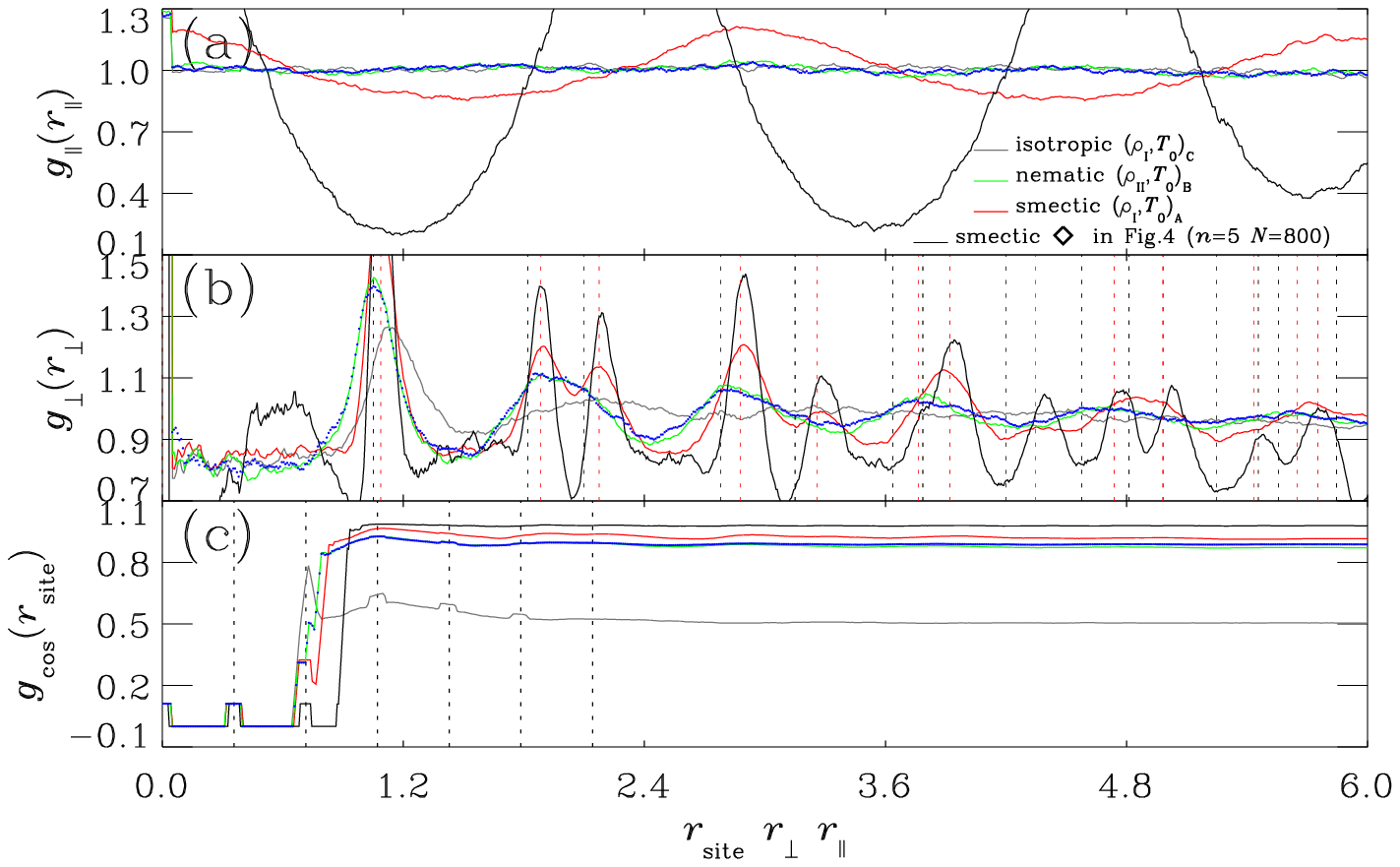} %g69655000.eps}
\end{center}
\end{figure}

%% Figure 7
\begin{figure}
%% \begin{figure}[tb]
\begin{center}
 (Here is Figure 7.)
 \includegraphics[width=0.255\textwidth]{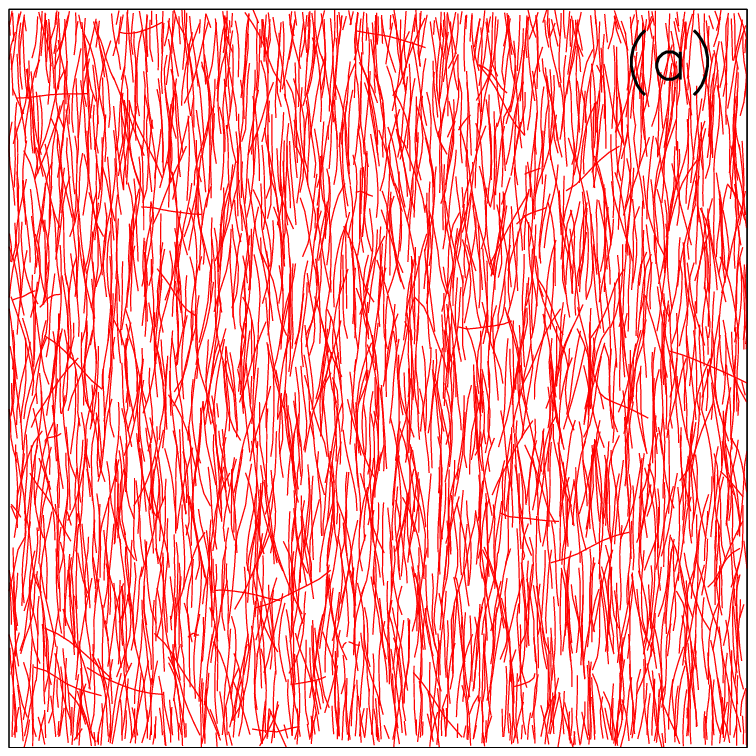}
 \includegraphics[width=0.255\textwidth]{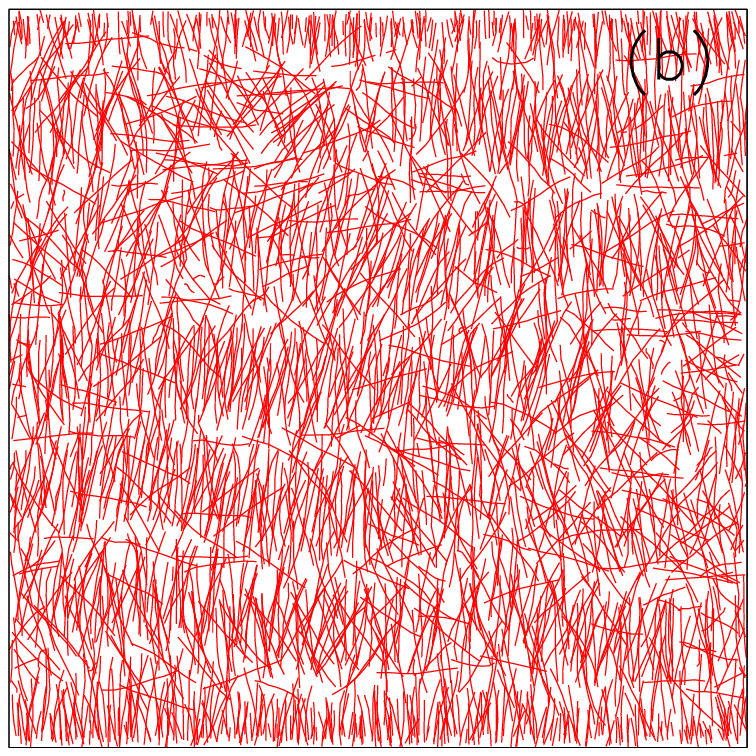}
\includegraphics[width=0.255\textwidth]{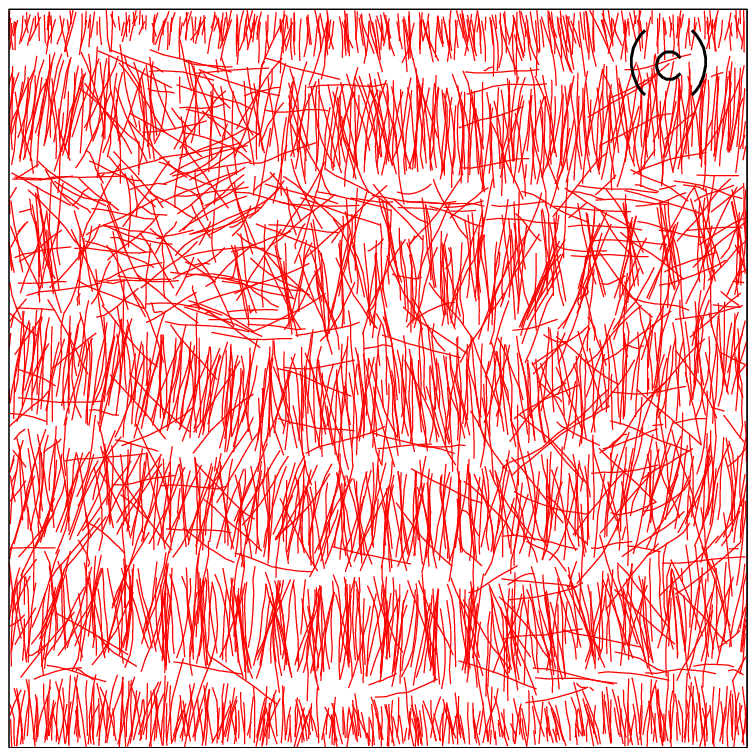}
\end{center}
\end{figure}

%% Figure 8
\begin{figure}
\begin{center}
 (Here is Figure 8.)
 \includegraphics[width=0.55\textwidth]{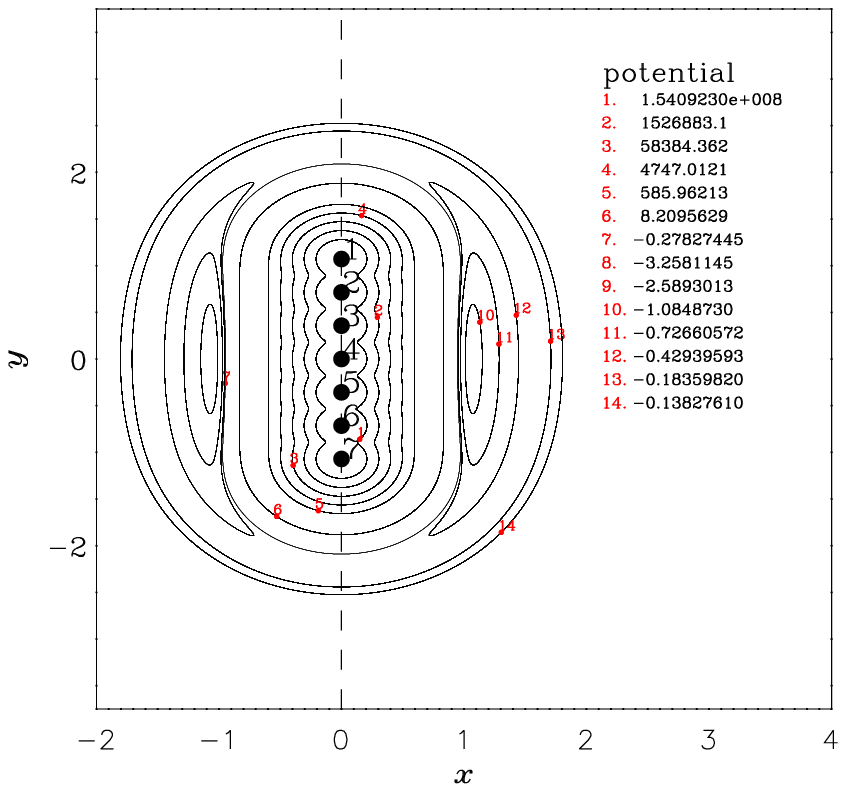}
\end{center}
\end{figure}

%% Figure 9
\begin{figure}
%% \begin{figure}[tb]
\begin{center}
(Here is Figure 9.)
 \includegraphics[width=0.465\textwidth]{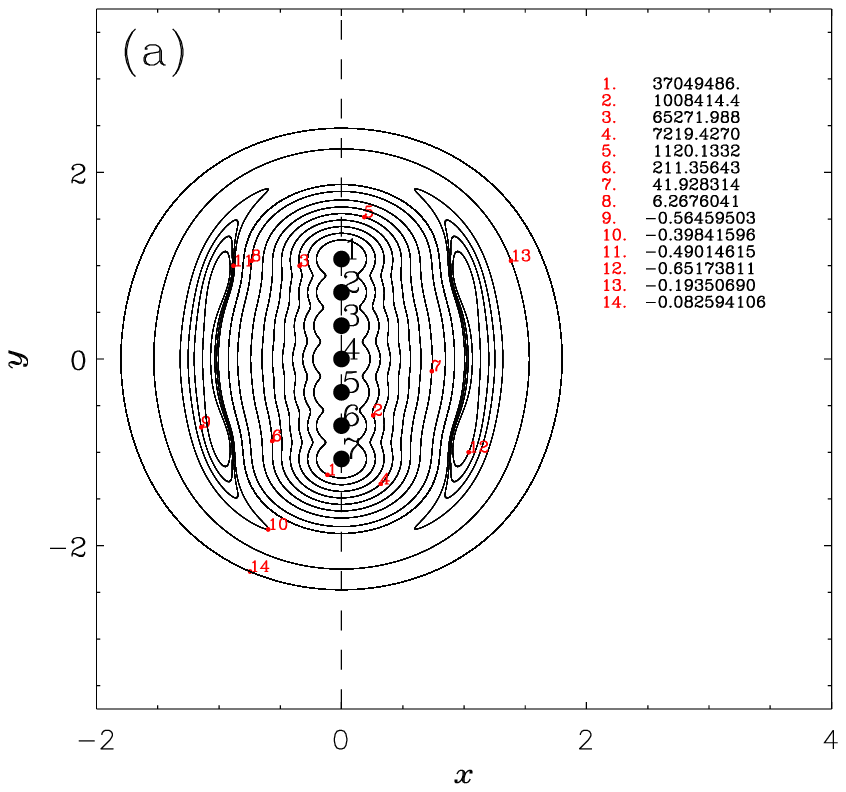} %g69655000.eps}
 \includegraphics[width=0.465\textwidth]{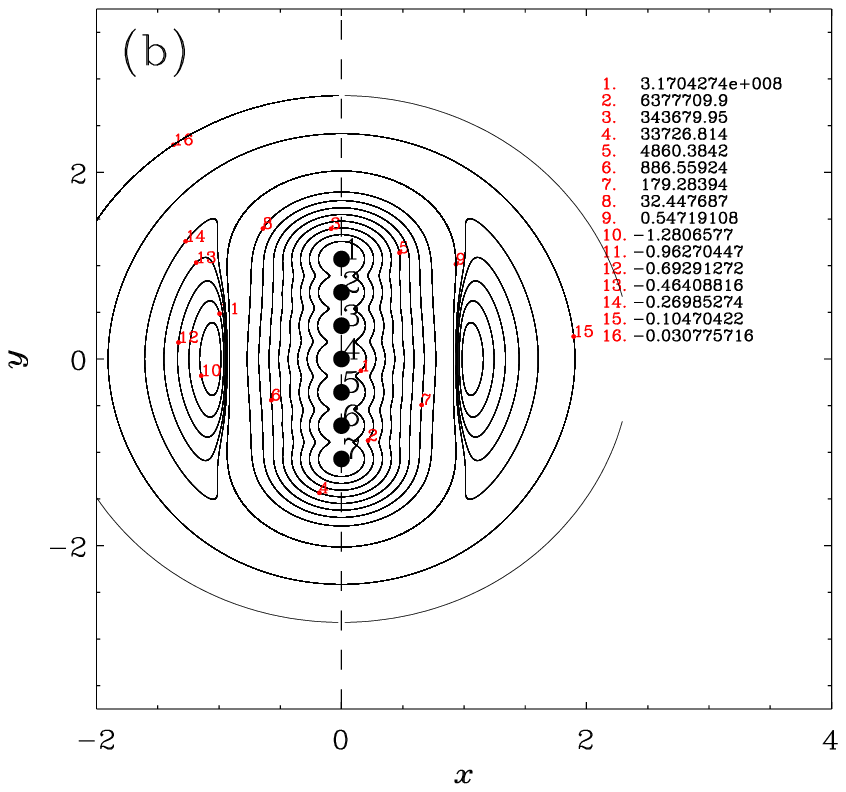} %g69655000.eps}
 \includegraphics[width=0.465\textwidth]{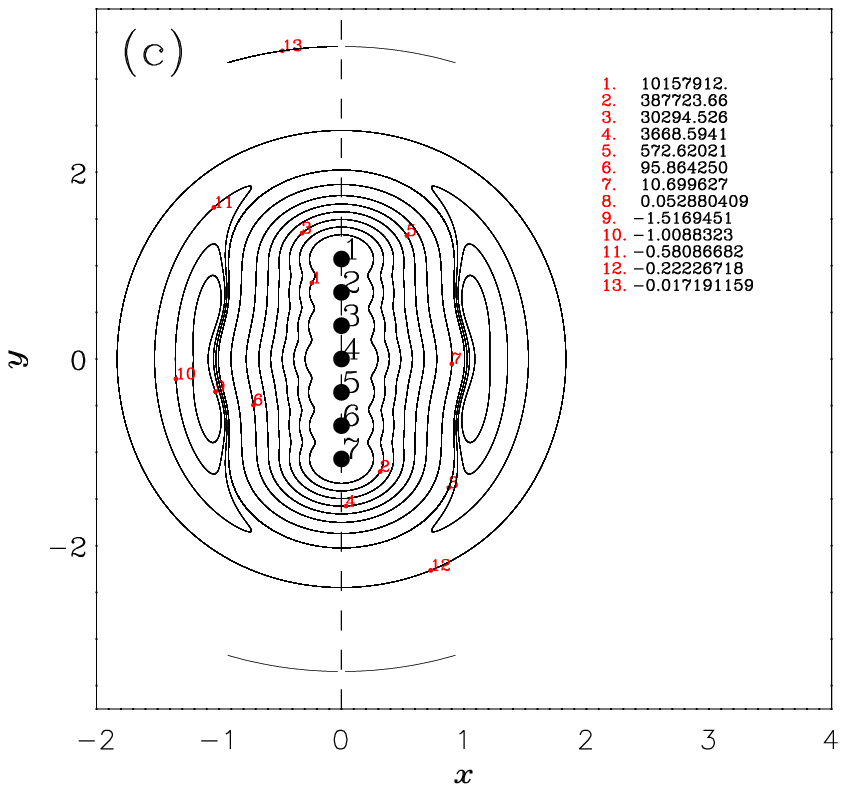} %g69655000.eps}
\end{center}
\end{figure}

\end{document}